\documentclass[pre,twocolumn,showpacs,amsmath,amssymb,nofootinbib,eqsecnum,floatfix]{revtex4}

\usepackage{graphicx}
\usepackage{bm}
\newtheorem{theorem}{Theorem}

\begin{document}

\title{Lyapunov exponents in constrained and unconstrained
ordinary differential equations}
\author{Michael D. Hartl}
\email{mhartl@tapir.caltech.edu}
\affiliation{
Department of Physics, California Institute of Technology, Pasadena CA 91125}
\date{May 5, 2003}
\begin{abstract}

We discuss several numerical
methods for calculating Lyapunov exponents (a quantitative
measure of chaos) in systems of ordinary differential equations.  We pay
particular attention to constrained systems, and we introduce a variety of
techniques to address the complications introduced by constraints. For all
cases considered, we develop both deviation vector methods, which follow the
time-evolution of the difference between two nearby trajectories, and Jacobian
methods, which use the Jacobian matrix to determine the true local behavior of
the system.  We also assess the merits of the various methods, and discuss
assorted subtleties and potential sources of error.

\end{abstract}
\maketitle

\section{Introduction}
\label{sec:introduction}

Chaos exists in a wide variety of nonlinear mathematical and physical systems,
and ordinary differential equations are no exception.  Since the original
discovery by Edward Lorenz of deterministic chaos in a toy atmosphere model
(consisting of twelve differential equations)~\cite{Lorenz},  a seemingly
endless variety of ODEs exhibiting extreme sensitivity to initial conditions
has emerged.  Many tools, both qualitative and quantitative, have been
developed to investigate this chaotic behavior.   Perhaps the most important
quantitative measure of chaos is the method of Lyapunov exponents, which
indicate the average rate of separation for nearby trajectories.  (See
\cite{PRECB,PREYI,GottMel2002,PRER,PREVA,BarrowLevin2003} for some recent
investigations into measures of chaos and their applications.) The present
paper is concerned with general methods for calculating these exponents in
arbitrary systems of ODEs.  We first review the techniques for calculating
Lyapunov exponents in unconstrained systems~\cite{Ott1993,ASY1997} (where each
coordinate represents a true degree of freedom), and then introduce several new
methods for calculating Lyapunov exponents in constrained systems (where
there are more coordinates than there are degrees of freedom). 

A defining characteristic of a chaotic dynamical system is
\emph{sensitive dependence on initial conditions}, and the Lyapunov exponents
are a way of quantifying this sensitivity.  In a system of ordinary
differential equations, 
this sensitive dependence corresponds to an exponential 
separation of nearby phase-space
trajectories: if two initial conditions are initially separated by a
distance~$\epsilon_0$, the total separation grows (on average) according to
\begin{equation}
\label{eq:Lyap}
\epsilon(t) = \epsilon_0\,e^{\lambda t},
\end{equation}
where~$\lambda$ is a positive constant (with units of inverse time) called the
\emph{Lyapunov exponent}.  Two important caveats to Eq.~(\ref{eq:Lyap}) are 
necessary.  First, this prescription yields only the \emph{largest} Lyapunov
exponent, but a dynamical system with~$n$ degrees of freedom has in general~$n$
such exponents.
Second, Eq.~(\ref{eq:Lyap}) does not constitute a rigorous definition, since it
defines a true Lyapunov exponent only if $\epsilon$
is ``infinitesimal.''  A more precise definition of Lyapunov exponents involves
the true local behavior of the dynamical system, i.e., the derivative or its
higher-dimensional generalization.

We can go beyond Eq.~(\ref{eq:Lyap}) to determine (at least in principle)
all~$n$ Lyapunov exponents by considering not just one nearby initial
condition, but rather a ball of initial conditions with radius $\epsilon_0$. 
As discussed in Sec.~\ref{sec:unconstrained}, this ball evolves into an
$n$-dimensional ellipsoid under the time-evolution of the flow, and the lengths
of this ellipsoid's principal axes determine the Lyapunov exponents.  We will
see that there are many advantages to this ellipsoid view, both conceptual and
computational.

We discuss in Secs.~\ref{sec:unconstrained} and~\ref{sec:constrained} several
techniques for calculating Lyapunov exponents in ODEs, and compare the relative
merits of the various methods.  We take special care to explain methods for the
calculation of all $n$~Lyapunov exponents. Our principal examples are two
well-studied and simple systems: the Lorenz equations (Sec.~\ref{sec:Lorenz})
and the forced damped pendulum (Sec.~\ref{sec:fdp}).  The techniques and code
were developed and tested on the much more complex problem of spinning bodies
orbiting rotating (Kerr) black holes, as discussed briefly in
Sec.~\ref{sec:constrained_ellipsoid_method} and at
length in~\cite{Hartl_2002_1,Hartl_2002_2}.

Our two model systems are unconstrained, so that each variable represents a
true degree of freedom. As we see in Sec.~\ref{sec:constrained}, following the
evolution of a phase-space ellipsoid---and hence calculating the Lyapunov
exponents---becomes problematic when the system is constrained. 
Such systems are common in physics, with constraints arising for both
mathematical and physical reasons.  For example, instead of using the
angle~$\theta$ to describe the position of a pendulum, we may find it
mathematically convenient to integrate the equations of motion in Cartesian
coordinates $(x, y)$, with a constraint on the value of $x^2+y^2$.  Another
example is a spinning astronomical body, whose spin is typically described by
the components of its spin vector $\mathbf{S} = (S_x, S_y, S_z)$.  On physical
grounds, we might wish to fix the magnitude~$\|\mathbf{S}\| = S =
\sqrt{S_x^2+S_y^2+S_z^2}$, so that only
two of the three spin components represent true degrees of freedom.

We describe in Sec.~\ref{sec:constrained} three methods for finding Lyapunov
exponents in constrained systems.  Our principal example of a constrained
system is the forced damped pendulum described in Cartesian coordinates, a
system chosen both for its conceptual simplicity and to facilitate comparison
with the same system without constraints.  We also show the application of
these techniques to the dynamics of spinning compact objects in general
relativity.  It was the investigation of these constrained systems
in~\cite{Hartl_2002_1} that led to the development of the key ideas described
in this paper.

We have developed a general-purpose implementation of the principal algorithms
in this paper in~C++, which is available for download~\cite{HartlSoftware}. 
The user must specify the system of equations (and a Jacobian matrix if
necessary), as well as a few other parameters, but the main procedures are not
tied to any particular system.  Most of the results in this paper were
calculated using this implementation.

We use boldface to indicate Euclidean vectors, and the symbol~$\log$ signifies
the natural logarithm $\log_e$ in all cases.  We refer to the principal
semiaxes of an $n$-dimensional ellipsoid as ``axes'' or ``principal axes'' for
brevity.

\section{Lyapunov exponents in unconstrained flows}
\label{sec:unconstrained}

There are two primary approaches to calculating Lyapunov exponents in systems
of ordinary differential equations.  The first method involves the integration
of two trajectories initially separated by a small deviation vector; we obtain
a measure of the divergence rate by  keeping track of the length of this
deviation vector.  We refer to this as the \emph{deviation vector method}. The
second method uses a rigorous linearization of the equations of motion (the
Jacobian matrix) in order to capture the true local behavior of the dynamical
system.   We call this the \emph{Jacobian method}. Though computationally
slower, the Jacobian method is more rigorous, and also opens the possibility of
calculating more than just the principal exponent.  In this section we discuss
these two methods, and several variations on each theme, in the context of
unconstrained dynamical systems.

\begin{figure}
\begin{center}
\includegraphics[width=3in]{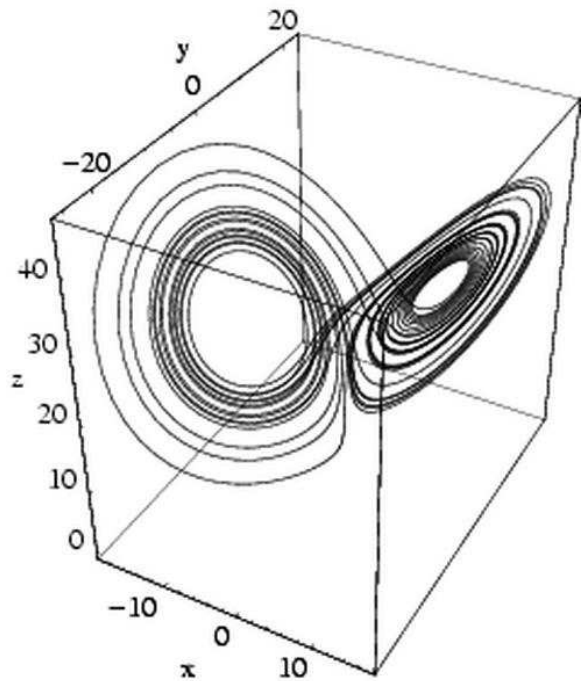}
\end{center}
\caption{\label{fig:Lorenz_attractor}
The Lorenz attractor.  All initial conditions except the origin (which is an
unstable equilibrium) are attracted to the figure shown.}
\end{figure}

When discussing Lyapunov exponents in ordinary differential equations, it is
valuable to have both a general abstract system and a specific concrete example
in mind.  Abstractly, we write the coordinates of the system as a single
$n$-dimensional vector~$\mathbf{y}$ that lives in the $n$-dimensional phase
space, and we write the equations of motion
as a system of first-order differential equations:
\begin{equation}
\label{eq:f}
\frac{d\mathbf{y}}{dt} = \mathbf{f}(\mathbf{y}).
\end{equation}
We will refer to a solution to Eq.~(\ref{eq:f}) as a \emph{flow}.
As a specific example, consider the Lorenz system of equations:
\begin{eqnarray}
\label{eq:Lorenz}
\dot{ x} & = & -\sigma x + \sigma y\nonumber\\
\dot{ y }&=& -xz+rx-y\\
\dot{ z}&=&xy-bz\nonumber,
\end{eqnarray}
where $\sigma$, $r$, and $b$ are constants.  In the notation of 
Eq.~(\ref{eq:f}), we then have $\mathbf{y} = (x, y, z)$ and
$\mathbf{f}(\mathbf{y}) = (-\sigma x + \sigma y, -xz+rx-y, xy-bz)$. 
The Lorenz equations exhibit chaos for a wide
variety of parameter values; in this paper, for simplicity we consider only one
such set: $\sigma = 10$, $b = 8/3$, and $r = 28$.  For these parameter values,
all initial conditions except the origin asymptote 
to the elegant Lorenz attractor (Fig.~\ref{fig:Lorenz_attractor}).

\subsection{The deviation vector method}
\label{sec:deviation_vector}

The most straightforward method for calculating the largest Lyapunov exponent
is to consider an initial point~$\mathbf{y}^{(1)}_0 = \mathbf{y}_0$ and a
nearby point $\mathbf{y}^{(2)}_0 = \mathbf{y}_0 + \delta\mathbf{y}_0$, and then
evolve both points forward, keeping track of the difference
$\delta\mathbf{y} \equiv \mathbf{y}^{(2)} - \mathbf{y}^{(1)}$. If the
motion is chaotic, then exponential separation implies that 
\begin{equation}
\|\delta\mathbf{y}\| = e^{\lambda_\mathrm{max} t}\,\|\delta\mathbf{y}_0\|,
\end{equation}
so that the largest exponent is
\begin{equation}
\label{eq:lambda_max}
\lambda_\mathrm{max} = 
\frac{\log{[r_e(t)]}}{t},
\end{equation}
where we write 
\begin{equation}
r_e= \|\delta\mathbf{y}\|/\|\delta\mathbf{y}_0\|,
\end{equation}
with a subscript~$e$ that anticipates the ellipsoid axis discussed in
Sec.~\ref{sec:ellipsoids}. Here $\|\cdot\|$ denotes the Euclidean norm (though
in principle any positive-definite norm will do~\cite{EckmannRuelle1985}).  It
is convenient to  display the results of this process graphically by plotting
$\log{[r_e(t)]}$ vs.~$t$, which we refer to as a \emph{Lyapunov plot}; since
Eq.~(\ref{eq:lambda_max}) is equivalent to $\log{[r_e(t)]} =
\lambda_\mathrm{max} t$, such plots should be approximately linear, with slope
equal to the principal Lyapunov exponent. (In practice, to extract the slope we
perform a least-squares fit to the simulation data, which is less sensitive to
fluctuations in the value of $\log{[r_e(t)]}$ than the ratio
$\log{[r_e(t_f)]}/t_f$ at the final time.) We
refer to this technique as the \emph{(unrescaled) deviation vector method}.

It is important to note that,  because of the problem of \emph{saturation},
Eq.~(\ref{eq:lambda_max}) does not define a true Lyapunov exponent.  In a
chaotic system, any deviation~$\delta\mathbf{y}_0$, no matter how small, will
eventually \emph{saturate}, i.e., it will grow so large that it no longer
represents the \emph{local} behavior of the dynamical system.  Moreover,
chaotic systems are bounded by definition [in order to eliminate trivial
exponential separation of the form $x(t) = x_0\,e^{\lambda t}$], so there is
some bound~$B$ on the distance between any two trajectories.  As a result, in
the infinite time limit Eq.~(\ref{eq:lambda_max}) gives
\begin{equation}
\lambda_\mathrm{max} =\lim_{t\rightarrow\infty} 
\frac{\log{\|\delta\mathbf{y}\|/\|\delta\mathbf{y}_0\|}}{t} \leq 
\lim_{t\rightarrow\infty}\frac{\log{B/\|\delta\mathbf{y}_0\|}}{t} = 0.
\end{equation}
In the na\"{\i}ve unrescaled deviation vector method,
the calculated exponent is always zero because of saturation.  

One solution to the saturation problem is to \emph{rescale} the deviation once
it grows too large.  For example, suppose that we set
$\|\delta\mathbf{y}_0\|=\epsilon$ for some small~$\epsilon$ (say $10^{-8}$), 
and then allow the deviation to grow by at most a factor of~$f$.  Then,
whenever  $\|\delta\mathbf{y}\|\geq f\,\|\delta\mathbf{y}_0\|$, we rescale the
deviation back to a size~$\epsilon$ and record the length $R_i =
\|\delta\mathbf{y}\|/\|\delta\mathbf{y}_0\|$ of the expanded vector.  If we
perform~$N$ such rescalings in the course of a calculation, the total expansion
of the initial vector is then
\begin{equation}
r_e = \frac{\|\delta\mathbf{y}_f\|}{\|\delta\mathbf{y}_0\|}\prod_{i=1}^N R_i,
\end{equation}
where $\delta\mathbf{y}_f$ is the final size of the (rescaled) 
separation vector.
Applying Eq.~(\ref{eq:lambda_max}), we see that the approximate Lyapunov 
exponent satisfies
\begin{equation}
\lambda_\mathrm{max} =
\frac{1}{t}
\left[\log{\left(\frac{\|\delta\mathbf{y}_f\|}{\|\delta\mathbf{y}_0\|}\right)}
+ \sum_{i=1}^N \log{R_i}\right].
\end{equation}
We refer to this as the \emph{(rescaled) deviation vector method}.

\begin{figure}
\begin{center}
\includegraphics[width=3in]{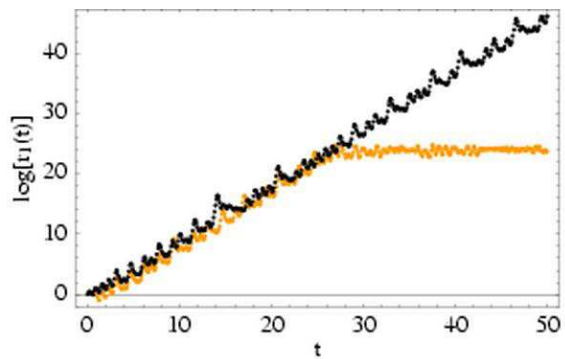}
\end{center}
\caption{\label{fig:deviation_compare}
Comparison of the unrescaled (light) and rescaled (dark) deviation vector
methods for calculating the principal Lyapunov exponent of the Lorenz system
[Eq.~(\ref{eq:Lorenz})].  The slope of the
rescaled line is the Lyapunov exponent ($\lambda_\mathrm{max} = 0.905\pm0.003$;
see Sec.~\ref{sec:Lorenz}). 
The initial deviation is $\|\delta\mathbf{y}_0\| = 10^{-8}$, and rescaling
occurs (for the rescaled method) if $\|\delta\mathbf{y}\| \geq 10^{-2}$.
Note the saturation
of the unrescaled approach once the deviation has grown too large.}
\end{figure}

The rescaled deviation vector method is not particularly robust compared to the
rigorous method described below (Sec.~\ref{sec:Jacobian_method}), and there are
significant complications when applying it to constrained systems, but if
implemented with care it provides a fast and accurate estimate for the largest
Lyapunov exponent. Fig.~\ref{fig:deviation_compare} shows both the rescaled and
unrescaled deviation vector methods applied to the Lorenz system
[Eq.~(\ref{eq:Lorenz})].  Note in particular the saturation of the unrescaled
approach.  We discuss the limitations of the rescaled method further in
Sec.~\ref{sec:comparing}.

\subsection{The Jacobian method}
\label{sec:Jacobian_method}
 
Although the deviation vector method suffices for practical calculation in many
cases, in essence it amounts to taking a numerical derivative.  For a
one-dimensional function of one variable, we can approximate the derivative
at~$x=x_0$ using
\begin{equation}
f'(x_0) \approx \frac{f(x_0+\epsilon) - f(x_0)}{\epsilon},
\end{equation}
for some $\epsilon\ll 1$, but this prescription is notoriously inaccurate as a
numerical calculation~\cite{NumRec}.  Of course, 
it is better (if possible) to calculate
the analytical derivative~$f'(x)$ and evaluate it at~$x_0$.  The
higher-dimensional generalization of this is the Jacobian matrix, which
describes
the local (linear) behavior of a higher-dimensional function.
In the context of
a dynamical system, this means that we can find the time-evolution of a small
deviation~$\delta\mathbf{y}$ using the rigorous linearization of the equations
of motion:
\begin{equation}
\label{eq:fydy}
\mathbf{f}(\mathbf{y}+\delta\mathbf{y})-\mathbf{f}(\mathbf{y})=\mathbf{Df}\cdot
\delta\mathbf{y}+O(\|\delta\mathbf{y}\|^2),
\end{equation}
where
\begin{equation}
\label{eq:jacobian_def}
(\mathbf{Df})_{ij}=\frac{\partial f_i}{\partial x^j}
\end{equation}
is the Jacobian matrix evaluated along the flow.  
For example, for the Lorenz system 
[Eq.~(\ref{eq:Lorenz})] we have
\begin{equation}
\label{eq:LorenzJacobian}
\mathbf{Df} = \left(
\begin{array}{ccc}
	-\sigma & \sigma & 0\\
	r-z(t) & -1 & -x(t)\\
	y(t) & x(t) & -b\\
\end{array}
\right),
\end{equation}
where we write the coordinates as functions of time to emphasize that
Eq.~(\ref{eq:LorenzJacobian}) is different at each time~$t$.

\subsubsection{Jacobian diagnostic}

One note about Jacobian matrices is worth mentioning: practical
experience has shown that errors occasionally creep into the calculations
leading to the Jacobian matrix, especially if the equations of motion are
complicated.  It is therefore worthwhile to note that Eq.~(\ref{eq:fydy})
provides an invaluable diagnostic: calculate the quantity
\begin{equation}
\label{eq:diagnostic}
\Delta = \mathbf{f}(\mathbf{y}+\delta\mathbf{y})-\mathbf{f}(\mathbf{y})-
\mathbf{Df}\cdot\delta\mathbf{y}
\end{equation}
for varying values of~$\|\delta\mathbf{y}\|$; if $\Delta$ does not 
generally scale as 
$\|\delta\mathbf{y}\|^2$, then something is amiss.  (The routines
in~\cite{HartlSoftware} include this important Jacobian diagnostic function.)

\subsubsection{The principal exponent}

The main value of Eq.~(\ref{eq:fydy}) in the context of a dynamical system is
its combination with Eq.~(\ref{eq:f}) to yield an equation of motion for the
deviation~$\delta\mathbf{y}$:
\begin{equation}
\mathbf{f}(\mathbf{y} + \delta\mathbf{y}) = 
\frac{d}{dt}(\mathbf{y} + \delta\mathbf{y}) = \mathbf{f}(\mathbf{y}) +
\frac{d(\delta\mathbf{y})}{dt},
\end{equation} 
so that (discarding terms higher than linear order) Eq.~(\ref{eq:fydy}) gives
\begin{equation}
\frac{d(\delta\mathbf{y})}{dt} = \mathbf{Df}\cdot\delta\mathbf{y}.
\end{equation}
This equation is only approximately true for finite (that is, non-infinitesimal)
deviations, but we can take the infinitesimal limit by identifying the deviation
$\delta\mathbf{y}$ with an element~$\bm{\xi}$ in the tangent space
at~$\mathbf{y}$.  This leads to an exact equation for~$\bm{\xi}$:
\begin{equation}
\label{eq:xi}
\frac{d\bm{\xi}}{dt} = \mathbf{Df}\cdot\bm{\xi}.
\end{equation}
The initial value of~$\bm{\xi}$ is arbitrary, but it is convenient to require
that $\|\bm{\xi}_0\|=1$, so that the factor by which~$\bm{\xi}$ has grown
at some later time~$t$ is simply~$\|\bm{\xi}(t)\|$.

\begin{figure}
\begin{center}
\includegraphics[width=3in]{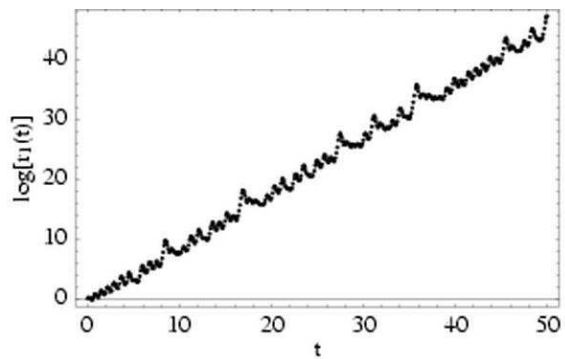}
\end{center}
\caption{\label{fig:Lorenz_Jacobian_1}
The natural logarithm of the tangent vector length $r_1\equiv\|\bm{\xi}(t)\|$
vs.~$t$ for the Lorenz system.  The slope of the
rescaled line is the system's largest Lyapunov exponent 
($\lambda_\mathrm{max}\approx0.905$).  
The figure
and exponent are virtually identical to the rescaled deviation method show in
Fig.~\ref{fig:deviation_compare}.}
\end{figure}

The core of the \emph{Jacobian method} for the principal Lyapunov exponent
is to solve Eqs.~(\ref{eq:f}) and~(\ref{eq:xi}) as a coupled set of
differential equations.  As in Sec.~\ref{sec:deviation_vector}, for chaotic
systems the length of the deviation vector will grow exponentially, so that
\begin{equation}
\|\bm{\xi}(t)\| \approx e^{\lambda_\mathrm{max}t},
\end{equation}
which implies that
\begin{equation}
\label{eq:xi_lambda}
\lambda_\mathrm{max} = \frac{\log{\|\bm{\xi}(t)\|}}{t}.
\end{equation}
For sufficiently large values of~$t$, 
Eq.~(\ref{eq:xi_lambda}) provides an approximation for the largest Lyapunov
exponent.  It is essential to understand that there is no restriction on the
length of the tangent  vector~$\bm{\xi}$: the Jacobian method \emph{does not
saturate}.  The only limitation on the size of~$\bm{\xi}$ in practice is the
maximum representable floating point number on the computer.

\begin{figure}
\begin{center}
\includegraphics[width=3in]{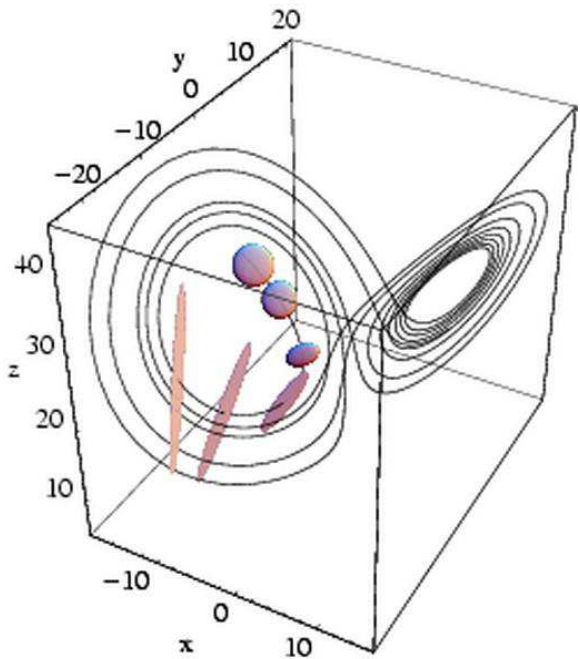}
\end{center}
\caption{\label{fig:Lorenz_ell}
The Lorenz system with an evolving ellipsoid.  The ellipsoid is calculated
exactly in the tangent space (for a total time~$t=0.4$) 
and is superposed on the phase space for the
purposes of visualization.  There is one expanding axis ($\sim e^{0.905\,t}$) 
and one contracting
axis ($\sim e^{-14.57\,t}$); 
the third axis has a fixed unit length (Sec.~\ref{sec:Lorenz}).}
\end{figure}

\subsubsection{Ellipsoids and multiple exponents}
\label{sec:ellipsoids}

\begin{figure}
\begin{center}
\includegraphics[width=3in]{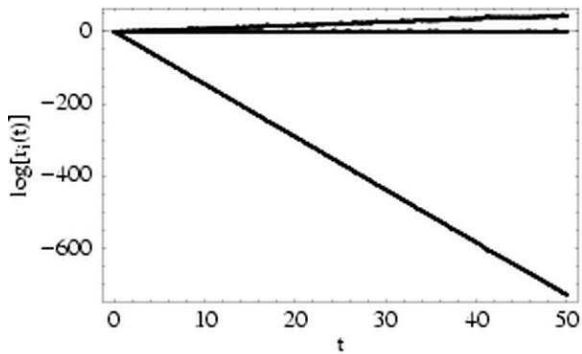}
\end{center}
\caption{\label{fig:Lorenz_Jacobian_3}
The natural logarithms of all three of the ellipsoid 
axes~$r_i$ vs.~$t$ for the Lorenz system, calculated using the Jacobian method
(Sec.~\ref{sec:Jacobian_method}).
The slopes are the Lyapunov exponents.  The three lines correspond to the
exponents $\lambda_1\approx0.905$, $\lambda_2\approx0.0$, and
$\lambda_3\approx -14.57$ (Sec.~\ref{sec:Lorenz}).  These values agree with the calculations
in~\cite{ASY1997}.}
\end{figure}

Although following the time-evolution of a tangent vector~$\bm{\xi}$ in place
of a finite deviation~$\delta\mathbf{y}$ solves the problem of saturation, it
still only allows us to determine the principal
exponent~$\lambda_\mathrm{max}$.  For a system with~$n$ degrees of freedom,
this leaves~$n-1$ exponents undetermined.  In order to calculate all~$n$
exponents, we must introduce~$n$ tangent vectors. (We discuss the value of
knowing all~$n$ exponents in Sec.~\ref{sec:multiple} below.) Since $n$ (linearly
independent) vectors span an $n$-dimensional ellipsoid, this leads to a
visualization of the Lyapunov exponents in terms of the evolution of a tangent
space ellipsoid (Fig.~\ref{fig:Lorenz_ell}).  Fig.~\ref{fig:Lorenz_Jacobian_3}
shows the corresponding Lyapunov plot.

The general method is to introduce a linearly independent set of  vectors
$\{\bm{\xi}^{(1)}, \bm{\xi}^{(2)}, \ldots, \bm{\xi}^{(n)}\}$.  It is convenient
to begin the integration with vectors that form the orthogonal axes of a unit
ball, so that the vectors  $\{\bm{\xi}^{(1)}_0, \bm{\xi}^{(2)}_0, \ldots,
\bm{\xi}^{(n)}_0\}$ are orthonormal. 
Each of  these tangent vectors satisfies its own version
of Eq.~(\ref{eq:xi}):
\begin{equation}
\label{eq:xi^n}
\frac{d\bm{\xi}^{(n)}}{dt} = \mathbf{Df}\cdot\bm{\xi}^{(n)}.
\end{equation}
If we combine the $n$~tangent vectors to form the columns of a 
matrix~$\mathbf{U}$, then Eq.~(\ref{eq:xi^n}) implies that
\begin{equation}
\label{eq:Udot}
\frac{d\mathbf{U}}{dt} = \mathbf{Df}\cdot\mathbf{U}.
\end{equation}
This equation, combined with Eq.~(\ref{eq:f}), describes the evolution of a
unit ball into an $n$-dimensional ellipsoid. 

The value of the tangent space ellipsoid is this: if $r_i$ is the $i$th
principal ellipsoid axis [and $r_i(0) = 1$], then 
\begin{equation}
\label{eq:L}
r_i(t) = e^{\lambda_i t},
\end{equation}
where $\lambda_i$ is the $i$th Lyapunov exponent.  That is, the ellipsoid's
axes
grow (or shrink) exponentially, and if $\lambda_i > 0$ for any~$i$ then the
system is chaotic~\cite{EckmannRuelle1985}. 
[Recall that we refer to the semiaxes as ``axes'' for
brevity (Sec.~\ref{sec:introduction}).] 
Turning Eq.~(\ref{eq:L}) around, we can find the $i$th
Lyapunov exponent by finding the average stretching (or shrinking)
per unit time of the $i$th principal ellipsoid axis:
\begin{equation}
\label{eq:lambda_i}
\lambda_i \approx \frac{\log{[r_i(t)]}}{t}.
\end{equation}
In practice, a more robust prescription is to record $\log{[r_i(t)]}$ as a
function of~$t$ and
perform a least-squares fit to the pairs $(t_j, \log{[r_i(t_j)]})$ 
to find the slope~$\lambda_i$.

Though Eq.~(\ref{eq:lambda_i}) provides an estimate for the $i$th Lyapunov
exponent, it requires us to find the $n$~principal axes of the final ellipsoid.
While it is true that 
the columns of the final matrix $\mathbf{U}_f$ necessarily span
an ellipsoid, but they are not in general orthogonal; in particular, 
the final tangent
vectors do not necessarily 
coincide with the ellipsoid's principal axes.  A first step in
extracting these axes is to note an important theorem in linear algebra
(see~\cite{ASY1997} for a proof):
\begin{theorem}
\label{thm:axes}
Let $A$ be an $n\times n$ real matrix consisting of $n$~linearly independent 
column
vectors $\{\mathbf{v}_i\}_{i=1}^n$, and 
let $\{s_i^2\}_{i=1}^n$ be the eigenvalues and $\{\mathbf{u}_i\}_{i=1}^n$ 
the normalized eigenvectors of~$A^{T} A$ (where $A^T$ is the  
transpose of~$A$).
Then $\{\mathbf{v}_i\}_{i=1}^n$ lie on an $n$-dimensional ellipsoid whose 
principal axes are $\{s_i\,\mathbf{u}_i\}_{i=1}^n$.
\end{theorem}
In other words, finding the principal axes of
the ellipsoid represented by a matrix~$A$ is equivalent to finding the
eigensystem of~$A^T A$. (We note that 
the ellipsoid is unique: any other matrix~$B$
whose columns $\{\mathbf{w}_i\}_{i=1}^n$ lie on the same ellipsoid as
$\{\mathbf{v}_i\}_{i=1}^n$ must necessarily give the same principal axes.)

In principle, we are done: simply evolve~$\mathbf{U}$ for a long time, and 
find the eigenvalues of $\mathbf{U}^T\mathbf{U}$.  In practice, this fails
miserably; every (generic) initial vector~$\bm{\xi}^{(i)}_0$ has some component
along the direction of greatest stretching, so \emph{all} initial tangent space
vectors eventually point approximately along the longest principal axis.  As a
result, all axes but the longest one are lost due to finite floating point
precision.  

The solution is to find new orthogonal axes as the system evolves.  In other
words, we can let the system evolve for some time~$T$, stop to calculate the
principal axes of the evolving ellipsoid, and then continue the integration. 
The method we advocate is the Gram-Schmidt orthogonalization procedure, which
results in an orthogonal set of vectors spanning the same volume as the
original ellipsoid, and with directions that converge to the true ellipsoid
axes.  This approach, originally described in~\cite{Ben1980}, is a common textbook
approach~\cite{ASY1997,Ott1993}, 
and was used successfully by the present author
in~\cite{Hartl_2002_1}.  Numerically, the Gram-Schmidt algorithm is subject to
considerable roundoff error~\cite{NumRec}, and is usually considered a poor
choice for orthogonalizing vectors, but in the context of dynamics its
performance has proven to be astonishingly robust.  (See
Sec.~\ref{sec:comparing} for
further discussion.)

We review briefly the Gram-Schmidt construction, and then indicate its use in
calculating Lyapunov exponents. Given~$n$ linearly-independent vectors $\{{\bf
u}_i\}$, the Gram-Schmidt procedure constructs $n$ orthogonal vectors $\{{\bf
v}_i\}$ that span the same space, given by
\begin{equation}
{\bf v}_i={\bf u}_i-\sum_{j=1}^{i-1}\frac{{\bf u}_i
\cdot{\bf v}_j}{\|{\bf v}_j\|^2}\,{\bf v}_j.
\end{equation}
To construct the $i$th orthogonal vector, we take the $i$th vector from the
original set and subtract off its projections onto the previous $i-1$ vectors
produced by the procedure.
The use of Gram-Schmidt in dynamics comes from observing that the resulting
vectors approximate the axes of the tangent space ellipsoid.  After the
first time~$T$, all of the vectors point mostly along the principal expanding
direction.  We may therefore pick any one as the first vector in the
Gram-Schmidt algorithm, so choose $\bm{\xi}_1\equiv{\bf u}_1$ without loss of
generality. If we let ${\bf e}_i$ denote unit vectors along the principal axes
and let $r_i$ be the lengths of those axes, the dynamics of the system
guarantees that the first vector ${\bf u}_1$ satisfies
\[
{\bf u}_1=r_1{\bf e}_1+r_2{\bf e}_2+
\cdots\approx r_1{\bf e}_1\equiv{\bf v}_1
\]
since ${\bf e}_1$ is the direction of fastest stretching. The second
vector ${\bf v}_2$ given by Gram-Schmidt is then 
\[
{\bf v}_2={\bf u}_1-\frac{{\bf u}_1\cdot{\bf v}_1}{\|{\bf v}_1\|^2}\,{\bf v}_1
\approx{\bf u}_1-r_1{\bf e}_1=r_2{\bf e}_2,
\]
with an error of order $r_2/r_1$. The procedure proceeds iteratively, with each
successive Gram-Schmidt step (approximately) subtracting off the contribution
due to the previous axis direction.  In principle, the system should be allowed
to expand to a point where $r_2\ll r_1$, but (amazingly) in practice
the Gram-Schmidt
procedure converges to accurate ellipsoid axes even when the system is
orthogonalized \emph{and even normalized}
on timescales short compared to the Lyapunov stretching
timescale.  As a result, the procedure below can be abused rather badly and
still give accurate results (Sec.~\ref{sec:comparing}).

\subsubsection{The algorithm in detail}
\label{sec:detail}

We summarize here the method used to calculate all the Lyapunov exponents of an
unconstrained dynamical system $\dot{\mathbf{y}} = \mathbf{f}(\mathbf{y})$
with~$n$ degrees of freedom:

\begin{enumerate}

\item Construct an orthonormal matrix~$\mathbf{U}_0$ whose columns
(the initial tangent vectors) span a unit ball,
and then integrate
\begin{equation}
\dot{\mathbf{y}} = \mathbf{f}(\mathbf{y})
\end{equation}
and
\begin{equation}
\dot{\mathbf{U}} = \mathbf{Df}\cdot\mathbf{U}
\end{equation}
as a coupled set of $2n$~differential equations.  
We recommend choosing a random
initial ball for genericity.

\item At various times~$t_j$, replace~$\mathbf{U}$ with the orthogonal axes of
the ellipsoid defined by~$\mathbf{U}$, using the Gram-Schmidt orthogonalization
procedure.  This can be done either every time~$T$, for some suitable choice
of~$T$, or every time the integrator takes a step.  We have found the latter
prescription to be especially robust in practice.

\item If the length of any axis exceeds some \emph{very} large value (say, near
the maximum representable floating point value), normalize the ellipsoid and
record the axis lengths
\begin{equation}
R_i^{(k)}\qquad\mbox{($i$th axis at $k$th rescaling)}
\end{equation}
at the rescaling time.  Do the same if any axis
is smaller than some very small number.  
%

\item Record the value of 
\begin{equation}
\label{eq:running_sum}
\log{r_i^{(j)}} = \log{[L_i(t_j)]} + \sum_{k=1}^{k_\mathrm{max}}\log{R_i^{(k)}}
\end{equation}
at each time~$t_j$, where $L_i$ is the $i$th principal axis length.  The second
term accounts for the axis lengths at the $k_\mathrm{max}$ rescaling times.  
Note
that if $t_j$ is a rescaling time itself, then $\log{[L_i(t_j)]} = \log{1} = 0$,
since by construction the ellipsoid has been normalized back to a unit ball.

\item After reaching the final number of time steps~$N$, perform a least
squares fit on the pairs $(t_j, \log{r_i^{(j)}})$ to find the
slopes~$\lambda_i$.  Since 
\begin{equation}
\log{[r_i(t)]\approx \lambda_i t},
\end{equation}
the slope $\lambda_i$ is the Lyapunov
exponent corresponding to the $i$th principal axis. Using the Gram-Schmidt
procedure should result in the relationship $\lambda_1>\ldots>\lambda_n$.

\end{enumerate}

\begin{figure}
\begin{center}
\includegraphics[width=3in]{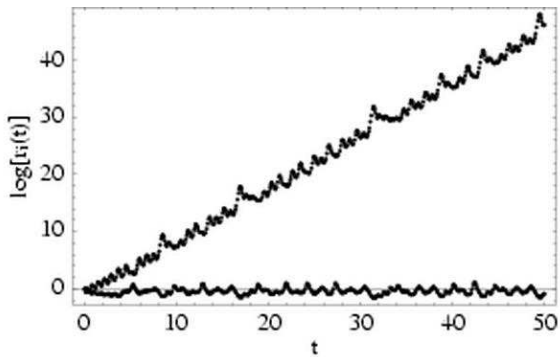}
\end{center}
\caption{\label{fig:Lorenz_Jacobian_2}
Close-up of Fig.~\ref{fig:Lorenz_Jacobian_3}, showing 
the natural logarithms of the two largest ellipsoid 
axes vs.~$t$ for the Lorenz system, calculated using the Jacobian method
(Sec.~\ref{sec:Jacobian_method}).
The slopes are the Lyapunov exponents.  The plot for the larger axis closely
matches the figures for the rescaled deviation vector method
(Fig.~\ref{fig:deviation_compare}) and the single tangent vector Jacobian method
(Fig.~\ref{fig:Lorenz_Jacobian_1}).}
\end{figure}

Most of the value of calculating $\lambda_i$ for $i>1$ comes from having
\emph{all}~$n$ of the exponents (Sec.~\ref{sec:multiple} below).  Nevertheless,
it is worth noting that the algorithm works for any value $0<m\leq n$, so the
method above can be used without alteration to find an arbitrary number of
exponents.  Fig.~\ref{fig:Lorenz_Jacobian_2} shows the axis growth for $m=2$ in
the Lorenz system, while Fig.~\ref{fig:Lorenz_Jacobian_3} shows the growth for
$m=n=3$.

\subsection{The value of multiple exponents}
\label{sec:multiple}

Calculating all the exponents of a system of differential equations allows us
to paint a more complete picture of the dynamics in several different
ways.  In particular, with all $n$~exponents comes the ability to visualize the
entire phase space ellipsoid (instead of just its principal axis), as in
Fig.~\ref{fig:Lorenz_ell}.  
Another important benefit of knowing all the exponents is a determination of 
dissipative
or conservative behavior.
Conservative flows preserve phase space volumes, while
dissipative flows contract volumes.  Geometrically, the volume~$V$ of an
ellipsoid is proportional to the product of its principal axes~$\{r_i\}$, so
that the ratio of the final to the initial volume is
\begin{equation}
\label{eq:V1}
\frac{V_f}{V_0} = \prod_i r_i,
\end{equation}
assuming that the initial volume is a unit ball.
For dissipative systems,
phase space volumes in general contract exponentially according to
\begin{equation}
\label{eq:V2}
\frac{V_f}{V_0} = e^{-\Lambda t},
\end{equation}
where~$\Lambda$ is a positive constant.
Combining Eq.~(\ref{eq:V1}) and Eq.~(\ref{eq:V2}) yields
\begin{equation}
\label{eq:lam_exp}
\Lambda = -\log{\left(\frac{V_f}{V_0}\right)} = 
    -\log{\left(\prod_i r_i\right)} = -\sum_i\log{r_i} = -\sum_i\lambda_i,
\end{equation}
where the $\lambda_i$ are the Lyapunov exponents.  In other words, \emph{the
phase space volume contraction constant~$\Lambda$ is equal to minus the sum of 
the Lyapunov exponents}.  

If the Lyapunov exponents sum to zero, then the contraction factor vanishes,
and volumes are conserved---i.e, the system is conservative.  The special case
of Hamiltonian systems is of particular interest, since the equations of motion
for  many mechanical systems can be derived from a Hamiltonian.  The
Hamiltonian property strongly constrains the Lyapunov exponents, which must
\emph{cancel pairwise}: to each exponent~$+\lambda$ there corresponds a second
exponent~$-\lambda$~\cite{EckmannRuelle1985}.  Several examples of
this~$\pm\lambda$ property of Hamiltonian systems appear below.

Having all the Lyapunov exponents also allows us to verify that there is at
least one vanishing exponent, corresponding to motion tangent to the flow,
which must be the case for any chaotic system.  (See Ref.~\cite{ASY1997} for a
proof.)  Since we have finite numerical precision, we do not expect to find any
exponent to be identically zero, but \emph{some} exponent should always be
close to zero.  A practical criterion for ``close to zero'' is to compute error
estimates for the  least-squares fits advocated in Sec.~\ref{sec:detail};  an
exponent is ``close to zero'' if it is zero to within the standard error of the
fit.  Applications of this method appear in Sec.~\ref{sec:Lorenz} and
Sec.~\ref{sec:fdp} below.  It is worth noting that the fitting errors are not
the dominant source of variance in calculating Lyapunov exponents; variations in
the initial conditions and initial deviation vectors contribute more to the
uncertainty than errors in the fits.  See Sec.~\ref{sec:Lorenz} for further
discussion.

One final note deserves mention: the statement that 
$\Lambda = -\sum_i\lambda_i$
is equivalent to a theorem due to Liouville~\cite{ASY1997}, which relates the
volume contraction to the trace of the Jacobian matrix:
\begin{equation}
\frac{V_f}{V_0} = \exp{\left(\int_0^t \mathrm{Tr}\,\mathbf{Df}(t)\,dt\right)},
\end{equation}
where again we assume that~$V_0$ corresponds to a unit ball.
If the trace of the Jacobian matrix happens to be time-independent, then this
yields
\begin{equation}
\label{eq:tind_trace}
\frac{V_f}{V_0} = 
\exp{\left[(\mathrm{Tr}\,\mathbf{Df})\,t\right]},\qquad\mbox{(time-independent
trace)}
\end{equation}
so that Eq.~(\ref{eq:lam_exp}) gives $\Lambda = -\mathrm{Tr}\,\mathbf{Df}$.
In this special case, we can perform a consistency check by verifying that
\begin{equation}
\label{eq:lambda_trace}
\sum_i\lambda_i = \mathrm{Tr}\,\mathbf{Df}.\qquad\mbox{(time-independent
trace)}
\end{equation}

\subsection{Examples}

\subsubsection{The Lorenz system}
\label{sec:Lorenz}

Following the phase space ellipsoid allows us to visualize the dynamics of the
Lorenz system in an unusual way.  Fig.~\ref{fig:Lorenz_ell} shows the Lorenz
attractor together with the phase space ellipsoid for a short amount of time
($t_f = 0.4$).  The initial ball is evolved using Eq.~\ref{eq:Udot}, so it
represents the true tangent space evolution, which is then superposed on the
Lorenz phase space $(x, y, z)$.  It is evident that the initial ball is
stretched in one direction and flattened in another, as well as rotated.  (As
we shall see, the third direction is neither stretched nor squeezed,
corresponding to the zero exponent discussed in Sec.~\ref{sec:multiple}.)

By recording natural logarithms of the ellipsoid axes as the system evolves, we
can obtain numerical estimates for the Lyapunov exponents, as discussed in
Sec.~\ref{sec:detail}.  A plot of $\log{[r_i(t)]}$ vs.~$t$ appears
in Fig.~\ref{fig:Lorenz_Jacobian_3} 
for a final time $t_f = 50$, with the slopes giving approximate values for the
exponents.  Using a $t_f = 5000$ integration for greater accuracy
yields the estimates 
\begin{eqnarray}
\label{eq:Lorenz_lambda}
\lambda_1 & = & 0.905\pm9\times10^{-6}\nonumber\\
\lambda_2 & = & 1.5\times10^{-6}\pm1.7\times10^{-6}\\
\lambda_3 & = & -14.57\pm9\times10^{-6}\nonumber
\end{eqnarray}
for the parameter values $\sigma = 10$, $b = 8/3$, and $r = 28$.  
The $\pm$~values are the standard errors on the least-squares fit of
$\log{[r_i(t)]}$ vs.~$t$.
One of the
exponents is close to zero (as required for a flow)
in the sense of Sec.~\ref{sec:multiple}: the error in the fit not small compared
to the exponent.  [In the case shown in Eq.~(\ref{eq:Lorenz_lambda}), 
the ``error'' is actually 
\emph{larger} than the exponent.]
The other two exponents are clearly nonzero, with the positive exponent
indicating chaos.  

As mentioned briefly in Sec.~\ref{sec:multiple}, the largest source of variance
in calculating Lyapunov exponents is variations in the initial conditions, not
errors in the least-squares fits used to determine the exponents.    We express
the exponents in the form 
\begin{equation}
\bar{\lambda}\pm\frac{\sigma}{\sqrt{N}},
\end{equation}
 where 
\begin{equation}
\bar{\lambda}=\frac{1}{N}\,\sum_{j=1}^{N}\lambda^{(j)}
\end{equation}
is the sample mean and 
\begin{equation}
\sigma = \sqrt{\frac{1}{N-1}\,
    \sum_{j=1}^{N}\left(\lambda^{(j)}-\bar{\lambda}\right)^2}
\end{equation}
is the standard deviation.
For the Lorenz system, using a final integration time of $t_f = 5000$
for  $N=50$~random initial balls [all centered on the same initial value of
$(x_0, y_0, z_0)$] gives
\begin{eqnarray}
\label{eq:Lorenz_lambda2}
\lambda_1 & = & 0.9053\pm4.1\times10^{-4}\nonumber\\
\lambda_2 & = & -4.5\times10^{-6}\pm7.6\times10^{-7}\\
\lambda_3 & = & -14.5720\pm4.1\times10{-4}\nonumber
\end{eqnarray}
The values of the error are much greater than the
standard errors associated with the least-squares fit for the slope for any one
trial.  As expected, it is evident that $\lambda_2$ is consistent with zero.

There is a strongly expanding direction and a very strongly contracting
direction in the Lorenz system, and the volume contraction constant~$\Lambda$
is large: $\Lambda = -\sum_i\lambda_i = 13.67$, so that after a time~$t=5000$
the volume is an astonishingly small $6.75\times10^{-29674}$.  This is despite
the exponential growth of the largest principal axis, which grows in this same
time to a length $1.52\times10^{1965}$; the volume nevertheless contracts,
since the smallest axis shrinks to
$4.44\times10^{-31639}$ in the same time. We note that the periodic
renormalization and reorthogonalization of the ellipsoid axes is absolutely
essential from a numerical perspective, since these axis lengths are far above
and below the floating point (double precision) limits of
$\mathtt{xmax}\approx\mathtt{xmin^{-1}}\approx10^{308}$ on a typical
IEEE-compliant machine~\cite{NumRec}.

The Lorenz system affords an additional check on the numerically determined
exponents: the trace of the Jacobian matrix [Eq.~(\ref{eq:LorenzJacobian})] 
is time-independent,
so the exponents should satisfy Eq.~(\ref{eq:lambda_trace}):
\begin{equation}
\sum_i\lambda_i = -13.67 \stackrel{?}{=} 
\mathrm{Tr}\,\mathbf{Df} = -(\sigma + 1 + b) = -\frac{41}{3} \approx -13.67.
\end{equation}
Eq.~(\ref{eq:lambda_trace}) is thus well-satisfied.

\subsubsection{The forced damped pendulum}
\label{sec:fdp}

We turn now to our second principal example of a chaotic dynamical system, the
forced damped pendulum (FDP).  
This is a standard pendulum with damping and periodic
forcing; written as a first-order ODE, our equations are as follows: 
\begin{eqnarray}
\label{eq:fdp}
\dot{\theta} & = & \omega\nonumber\\
\dot{\omega} &=& -c\,\omega-\sin\theta+\rho\,\sin t\\
\dot{ t}&=&1\nonumber
\end{eqnarray}
Here~$c$ is the damping coefficient and~$\rho$ is the forcing amplitude, and
the gravitational acceleration~$g$ and pendulum length~$\ell$ are set to one
for simplicity.  We include the equation $\dot{ t} = 1$ so that the system is
autonomous (i.e., we remove the explicit time-dependence by treating time as a
dynamical variable with unit time derivative).  In addition to being an example
with transparent physical relevance (in contrast to the Lorenz system), the
forced damped pendulum, in slightly altered form, serves as a model constrained
system in Sec.~\ref{sec:constrained} below.

\begin{figure}
\begin{center}
\includegraphics[width=3in]{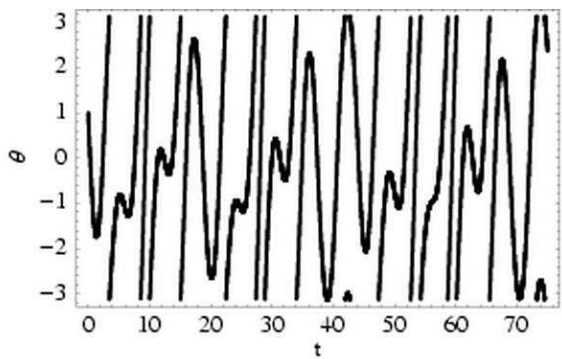}
\end{center}
\caption{\label{fig:theta_time}
$\theta$ vs.~$t$ for the forced damped pendulum [Eq.~(\ref{eq:fdp})].}
\end{figure}

\begin{figure}
\begin{center}
\includegraphics[width=3in]{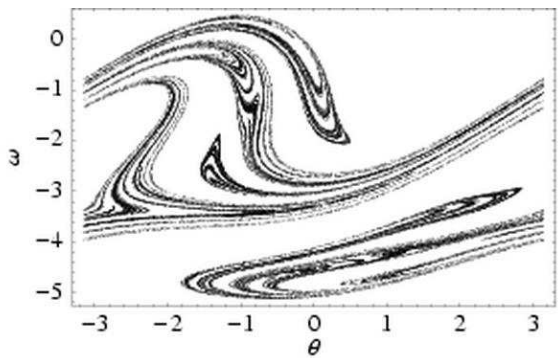}
\end{center}
\caption{\label{fig:time_2Pi}
$\omega$ vs.~$\theta$:
the time-$2\pi$ stroboscopic map for the forced damped pendulum.  A point
$(\omega=\dot{\theta},\theta)$ is plotted every time~$2\pi$, 
resulting in a fractal
attractor characteristic of dissipative chaos.}
\end{figure}

The forced damped pendulum is chaotic for many values of~$c$ and~$\rho$.  For
simplicity, in the present case we fix $c=0.1$ and $\rho = 2.5$.  A plot of
$\theta$ vs.~$t$ shows the system's erratic behavior
(Fig.~\ref{fig:theta_time}), but a more compelling picture of the dynamics
comes from a time-$2\pi$ stroboscopic map.  A time-$T$ map involves taking a
snapshot of the system every time~$T$ and then plotting $\omega$ vs.~$\theta$.
Since the forcing term in Eq.~(\ref{eq:fdp}) is $2\pi$-periodic, this provides
a natural value for~$T$ in the present case.  The resulting plot shows
the characteristic folding and stretching of a fractal attractor
(Fig.~\ref{fig:time_2Pi}), which for the FDP attracts almost all initial
conditions~\cite{ASY1997}.

\begin{figure}
\begin{center}
\includegraphics[width=3in]{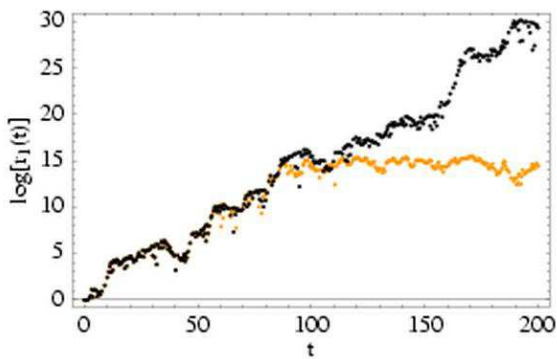}
\end{center}
\caption{\label{fig:Lyapunov_fdp}
The natural logarithms of all three of the ellipsoid 
axes~$r_i$ vs.~$t$ for the forced damped pendulum, 
calculated using the Jacobian method
(Sec.~\ref{sec:Jacobian_method}).
The slopes are the Lyapunov exponents.  The three lines correspond to the
exponents $\lambda_1 = 0.160\pm0.0049$, $\lambda_2 = 0.0$, and
$\lambda_3 = -0.262\pm0.0053$ (Sec.~\ref{sec:comparing} and
Table~\ref{table:compare}).}
\end{figure}

The forced damped pendulum is dissipative and strongly
chaotic.  We calculate the Lyapunov exponents (Fig.~\ref{fig:Lyapunov_fdp})
using the Jacobian matrix:
\begin{equation}
\label{eq:fdp_Jacobian}
\mathbf{Df} = \left(
\begin{array}{ccc}
	0 & 1 & 0\\
	-\cos\theta &~~-c~~& \rho\cos t\\
	0 & 0 & 0\\
\end{array}
\right),
\end{equation}
The Lyapunov exponents are (for a $t_f = 5\times10^4$ integration)
\begin{eqnarray}
\label{eq:fdp_lambda}
\lambda_1 & = & 0.160\pm7\times10^{-6}\nonumber\\
\lambda_2 & = & 8\times10^{-8}\pm1\times10^{-7}\\
\lambda_3 & = & -0.262\pm7\times10^{-6}\nonumber
\end{eqnarray}
where the error terms are the standard errors in the least-squares
fit for the slope. (See Sec.~\ref{sec:comparing} and especially
Table~\ref{table:compare} for the true errors due to varying initial
deviations.) One exponent
is consistent with zero (as required for a flow) 
to within the error of the fit.
The dissipation constant is $\Lambda = -\sum_i\lambda_i = 0.1$.   
The trace of the Jacobian matrix is
time-independent, so that $\mathrm{Tr}\,\mathbf{Df} = -c$, and indeed
$\sum_i\lambda_i = -0.1 = -c = \mathrm{Tr}\,\mathbf{Df}$ as predicted by
Eq.~\ref{eq:lambda_trace}.

The zero exponent in the FDP is associated with the time ``degree of freedom''
in the Jacobian: if we delete the final row and column of the Jacobian matrix,
only the positive and negative exponents remain (see, e.g.,
Fig.~\ref{fig:restricted_jacobian_both} below).  Since the time is not an
actual dynamical variable, for the remainder of this paper we will suppress
this ``time piece,'' but it is important to note that the time dependence is
absolutely crucial to the presence of chaos.  According to the
\emph{Poincar\'{e}-Bendixon theorem}~\cite{ASY1997}, an autonomous system of
differential equations with fewer than three degrees of freedom \emph{cannot}
be chaotic.  We will treat the FDP system as a time-dependent system with two
degrees of freedom, but the extra equation $\dot t = 1$ in the autonomous
formulation is what creates the potential for chaos.

\begin{figure}
\begin{center}
\includegraphics[width=3in]{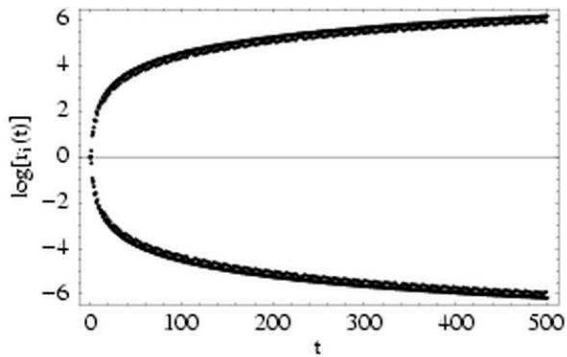}
\end{center}
\caption{\label{fig:Lyapunov_pendulum}
The natural logarithms of the ellipsoid 
axes~$r_i$ vs.~$t$ for the forced damped pendulum in the limit of zero
dissipation and zero forcing (i.e., a simple pendulum).
The Lyapunov exponents are zero, and the distance between nearby trajectories
grows linearly (leading to logarithmic growth in this log plot).  Nevertheless,
the Hamiltonian character of the system is manifest in the $\pm\lambda$
symmetry: for each exponent $+\lambda$, there is a corresponding
exponent~$-\lambda$.  In the nonchaotic limiting case shown here, 
the Lyapunov exponents approach zero symmetrically.}
\end{figure}

An instructive case to consider is the limit $c = \rho = 0$. In this limit, the
system is a simple pendulum, which is a Hamiltonian system.  A simple pendulum
is not chaotic, of course, and both its Lyapunov exponents are zero, but the
Hamiltonian character of the system nevertheless shows up in the $\pm\lambda$
property discussed above (Sec.~\ref{sec:multiple}): numerically, the exponents
approach zero in a symmetric fashion, as shown in
Fig.~\ref{fig:Lyapunov_pendulum}.

\section{Lyapunov exponents in constrained flows}
\label{sec:constrained}

We come now to the \emph{raison d'\^{e}tre} of this paper, namely, the
calculation of Lyapunov exponents for constrained systems. For pedagogical
purposes, our primary example is the forced damped pendulum with the position
written  in Cartesian coordinates. In addition to this instructive example, we
also discuss two constrained systems of astrophysical interest, involving the
orbits of spinning compact objects such as neutron stars or black holes (see,
e.g.,~\cite{Hartl_2002_1} and~\cite{Hartl_2002_2} and references therein).

Written in terms of the Cartesian coordinates
$(x, y) = (\cos\theta, \sin\theta)$, the equations of motion for the FDP 
[Eq.~(\ref{eq:fdp})] become (upon suppressing the time piece)
\begin{eqnarray}
\label{eq:fdpCartesian}
\dot{ x} & = & -\omega y\nonumber\\
\dot{ y }& = & \omega x\\
\dot{\omega} &=& -c\,\omega-y+\rho\,\sin t\nonumber
\end{eqnarray}
For a pendulum with unit radius, the Cartesian coordinates of the pendulum 
satisfy the
constraint
\begin{equation}
x^2 + y^2 = 1.
\end{equation}
Although it is certainly possible to use $(\dot{x}, \dot{y})$ in the equations
of motion, along with $(x, y)$, this is an unnecessary complication; in order
to keep the equations as simple as possible, we retain the  variable~$\omega$
in the equations of motion.

Developing the techniques for solving constrained systems using this toy
example has several advantages.  The equations of motion and the constraint are
extremely simple, which makes it easy to see the differences between the
constrained and unconstrained cases.  In addition, the constraint is easy to
visualize, and yet it captures the key properties of much more complicated
constraints.  Finally, since we have already solved the same problem in
unconstrained form, it is easy to verify that the techniques of this section
reproduce the results from Sec.~\ref{sec:fdp}.

\subsection{Constraint complications}
\label{sec:complications}

To see how constraints complicate the calculation of Lyapunov exponents,
consider an implementation of the deviation vector approach
(Sec.~\ref{sec:deviation_vector}). In the unconstrained forced damped pendulum,
given an initial condition, we would construct a deviated trajectory separated
by a small angle~$\delta\theta$ (and a small velocity~$\delta\omega$).  In the
constrained version, a na\"{\i}ve implementation would use a deviated
trajectory with spatial coordinates $x+\delta x$ and $y + \delta y$, where
$\delta\mathbf{y} = (\delta x, \delta y)$ is a small but otherwise arbitrary
deviation vector.  But the deviations are not independent; the deviated initial
condition must satisfy the constraint:
\begin{equation}
(x+\delta x)^2+(y+\delta y)^2 = 1.
\end{equation}
To lowest order in $\delta x$, we must have $\delta y = -(x/y)\,\delta x$.

We can now consider a more general case. Suppose there are $k$~constraints,
which we write as a $k$-dimensional vector equation
$\mathbf{C}(\mathbf{y})=\mathbf{0}$. (In our example, $\mathbf{C}$ has only one
component: with  $\mathbf{y} = (x, y, \omega)$, we have $C_1(\mathbf{y}) =
x^2+y^2-1=0$.)   Then if a point $\mathbf{y}$ satisfies the constraints, the
deviated trajectory must satisfy them as well:
\begin{equation}
\mathbf{C}(\mathbf{y}+\delta\mathbf{y})=\mathbf{0}.
\end{equation}
We will refer such a~$\delta\mathbf{y}$ as a \emph{constraint-satisfying
deviation}.

Let us outline one possible method for constructing such a
constraint-satisfying deviation.  Let~$n$ be the number of phase space
coordinates ($n=3$ for the constrained forced damped pendulum model). Consider
an $n$-dimensional vector~$\tilde{\mathbf{y}}_0$ that has~$d$ nonzero entries,
where~$d$ represents the true number of degrees of freedom ($d=2$ for the
constrained FDP).  Assume that we have some method for constructing
from~$\tilde{\mathbf{y}}_0$ an $n$-dimensional initial
condition~$\mathbf{y}_0$ that satisfies the constraints.  For example, we could
specify the initial values of~$x$ and~$\omega$, and then derive an initial
value of~$y$ using $y = \sqrt{1-x^2}$ (or $y = -\sqrt{1-x^2}$; more on this
later).  Now consider an $n$-dimensional vector $\tilde{\mathbf{y}}'_0 =
\tilde{\mathbf{y}}_0 + \delta\tilde{\mathbf{y}}_0$, which adds \emph{arbitrary}
deviations to~$d$ degrees of freedom.  We can then use the same method as above
to find~$\mathbf{y}'_0$ from $\tilde{\mathbf{y}}'_0$, and then set
\begin{equation}
\label{eq:constraint_dev}
\delta\mathbf{y}_0 = \mathbf{y}'_0 - \mathbf{y}_0
\end{equation}
to arrive at a constraint-satisfying deviation.

\begin{figure}
\begin{center}
\includegraphics[width=3in]{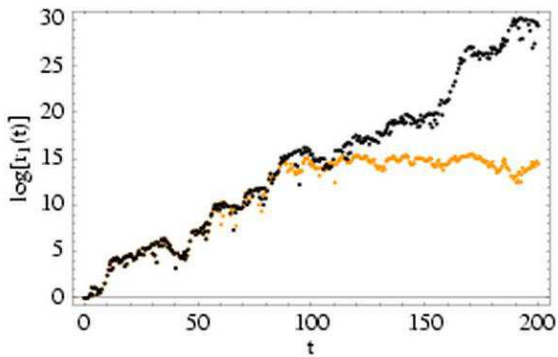}
\end{center}
\caption{\label{fig:constrained_deviation_compare}
Comparison of the unrescaled (light) and rescaled (dark) constrained
deviation vector
methods for calculating the principal Lyapunov exponent of the constrained
forced damped pendulum (Sec.~\ref{sec:rescaling_constrained}).  
The slope of the
rescaled line is the Lyapunov exponent, $\lambda_1 = 0.161\pm0.0046$
(Sec.~\ref{sec:comparing}).
The initial deviation is $\|\delta\mathbf{y}_0\| = 10^{-6}$, and rescaling
occurs (for the rescaled method) if $\|\delta\mathbf{y}\| \geq 10^{-2}$, which
happens~4 times in this figure.
As in Fig.~\ref{fig:deviation_compare}, the 
unrescaled approach saturates once the deviation has grown too large.}
\end{figure}

\subsection{Constrained deviation vectors}
\label{sec:constrained_deviation}

Having determined~$\delta\mathbf{y}_0$ by Eq.~(\ref{eq:constraint_dev}) (or by
some other method), we can immediately apply the unrescaled deviation
vector approach: simply track $\mathbf{y}'$ and $\mathbf{y}$ as the two
trajectories evolve, and monitor the length of $\delta\mathbf{y} =
\mathbf{y}' - \mathbf{y}$.  Since the equations of motion preserve the
constraint, the resulting~$\delta\mathbf{y}$ is always constraint-satisfying. 
The only subtlety is using a restricted norm to eliminate the extra degrees of
freedom; for example, the restricted FDP norm is
\begin{equation}
\|\delta\mathbf{y}\|_r = \sqrt{\delta x^2 + \delta\omega^2}
\end{equation}
if we choose to eliminate the~$y$ degree of freedom.  Since $\delta y \approx
-(x/y)\,\delta x$, using the full Euclidean distance would add the term
$\delta y^2 = (x^2/y^2)\,\delta x^2$ to the
expression under the square root, leading
to an overestimate for the principal exponent.  The
restricted norm avoids this problem by considering only true degrees of freedom.

\subsubsection{Rescaling for constrained systems}
\label{sec:rescaling_constrained}

In contrast to the simplicity of the unrescaled method, the \emph{rescaled}
deviation vector method requires great care, since a carelessly rescaled
deviation is not constraint-satisfying:
$\mathbf{C}(\mathbf{y}+\delta\mathbf{y}/r)\neq\mathbf{0}$ for a rescaling
factor~$r\neq 1$.  In this case, it is necessary to
extract~$\delta\tilde{\mathbf{y}}$ from $\delta\mathbf{y}$ and then rescale it
back to its initial size~$\|\delta\tilde{\mathbf{y}}_0\|$ using the restricted
norm. By reapplying the procedure leading to Eq.~(\ref{eq:constraint_dev}), we
then find a new (rescaled) constraint-satisfying~$\delta\mathbf{y}$ that
satisfies $\|\delta\mathbf{y}\|_r = \|\delta\tilde{\mathbf{y}}_0\|$. In this
case, \emph{it is essential that the new deviation vector have the same
constraint branches as the old one}.  For example, suppose that in the FDP case
the value of~$y$ is negative before the rescaling.  When calculating a
new~$\mathbf{y}'$ to arrive at the rescaled deviation~$\delta\mathbf{y}$, it is
then essential to choose the negative branch in the equation $y' =
\pm\sqrt{1-x'^2}$.  The result of implementing this constrained deviation vector
method to the forced damped pendulum appears in
Fig.~\ref{fig:constrained_deviation_compare}.

\begin{figure}
\begin{center}
\includegraphics[width=3in]{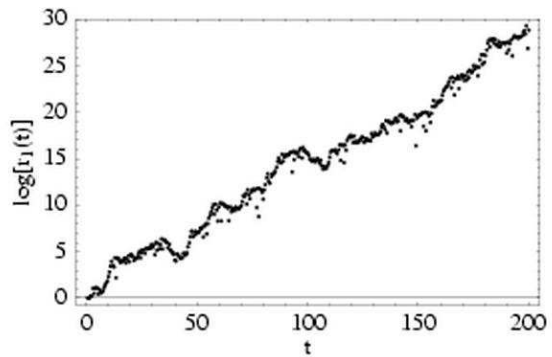}
\end{center}
\caption{\label{fig:restricted_jacobian}
The natural logarithm of the tangent vector length 
$r_1\equiv\|\bm{\xi}(t)\|_r$
vs.~$t$ for the constrained forced damped pendulum, 
using a constraint-satisfying tangent vector
(Sec.~\ref{sec:one_vector_constrained}).  We use the restricted norm
$\|\cdot\|_r$ to calculate phase space distances (see text).  Compare to
Fig.~\ref{fig:Lyapunov_fdp} 
(unconstrained Jacobian method) and
Fig.~\ref{fig:constrained_deviation_compare} 
(constrained deviation vector method).}
\end{figure}

\subsubsection{A Jacobian method for the largest exponent}
\label{sec:one_vector_constrained}

The method outlined above for unrescaled deviation vectors leads to a remarkably
simple implementation of the single tangent vector Jacobian method.  Given a
constraint-satisfying deviation~$\delta\mathbf{y}_0$, set
\begin{equation}
\bm{\xi}_0 = \delta\mathbf{y}_0/\|\delta\mathbf{y}_0\|_r,
\end{equation}
where $\|\cdot\|_r$ is a restricted norm on the~$d$ true degrees of freedom. 
We refer to such a~$\bm{\xi}$ as a \emph{constraint-satisfying tangent vector}. Since the equations of
motion preserve the constraints, we can evolve this tangent vector using
Eq.~(\ref{eq:xi}). The Jacobian method does not saturate, so we need only
rescale if~$\|\bm{\xi}\|_r$ approaches the floating point limit of the
computer.  We can then use a procedure based on the rescaled deviation method
to find a new (rescaled) constraint-satisfying tangent vector, but this is
typically unnecessary since by the time the floating point limit has been
reached we already have a good estimate of the principal Lyapunov exponent.  The
resulting Lyapunov plot for the constrained FDP appears in
Fig.~\ref{fig:restricted_jacobian}.

\subsubsection{Ellipsoid constraint complications}

We now have three methods at our disposal for calculating the \emph{largest}
Lyapunov exponent, but for~$d$ degrees of freedom there are~$d$ exponents. 
What of these other exponents? Here we find an essential difficulty in
implementing the ellipsoid method described in Sec.~\ref{sec:ellipsoids}.  The
core problem is this: the tangent vectors must be orthogonalized in order to
extract all~$d$ principal ellipsoid axes, but at the same time each tangent
vector must be constraint-satisfying.  Simply put, it is impossible in general
to satisfy the requirements of orthogonality and constraint satisfaction
simultaneously.  

We present here two different solutions to this problem, which we will refer to
as the restricted Jacobian method and the constrained ellipsoid method.

\subsection{Restricted Jacobian method}
\label{sec:restricted_Jacobian_method}

The most natural response to a system with more coordinates~$n$ than degrees of
freedom~$d$ is to eliminate the spurious degrees of freedom using the
constraints.  Unfortunately, this procedure is often difficult in practice:
solving the constraint equations may involve polynomial or transcendental
equations that have no simple closed form.  Even for the simple case of the
FDP, the sign ambiguity in $y = \pm\sqrt{1-x^2}$ makes a simple variable
substitution impractical.  Fortunately, such substitutions are unnecessary:
since the equations of motion preserve the constraints, there is no need in
general to eliminate $n-d$ coordinates.  In fact, constraints can be a virtue,
since they can be used to check the accuracy of the integration.

The same cannot be said of the Jacobian matrix.  As argued above, the extra
degrees of freedom lead to fundamental difficulties in applying the Jacobian
method for finding Lyapunov exponents; constraints, far from being a virtue,
are a considerable complication.  In contrast to the equations of motion,
though, it is relatively straightforward to eliminate the spurious degrees of
freedom.  The trick is to write a \emph{restricted} $d\times d$~Jacobian
matrix, with entries only for~$d$ coordinates.

An example should make this clear.  For the FDP system in constrained form, we
wish to eliminate one degree of freedom in the Jacobian matrix, and we can
choose to eliminate either~$x$ or~$y$.  Choosing the latter, the Jacobian
becomes
\begin{equation}
\mathbf{Df} = \left(
\begin{array}{cc}
	\frac{\partial\dot{ x}}{\partial x} & 
        \frac{\partial\dot{ x}}{\partial\omega}\\
	\frac{\partial\dot{\omega}}{\partial x} & 
        \frac{\partial\dot{ \omega}}{\partial\omega}\\
\end{array}
\right),
\end{equation}
where we have suppressed the derivatives with respect to the ``time degree of
freedom'' (as discussed in Sec.~\ref{sec:fdp}).  The term to focus on here is
$\partial\dot{ x}/\partial x$, which seems to be zero \emph{a priori} since 
$\dot{x} = -\omega y$, but
this is only true if we treat $x$ and $y$ as independent. Since we are eliminating
the~$y$ degree of freedom, we \emph{cannot} treat them as
independent; $y$ has a nonzero
derivative with respect to~$x$, so that
\begin{equation}
\frac{\partial\dot{ x}}{\partial x} = -\omega\frac{\partial y}{\partial x}.
\end{equation}

\begin{figure}
\begin{center}
\includegraphics[width=3in]{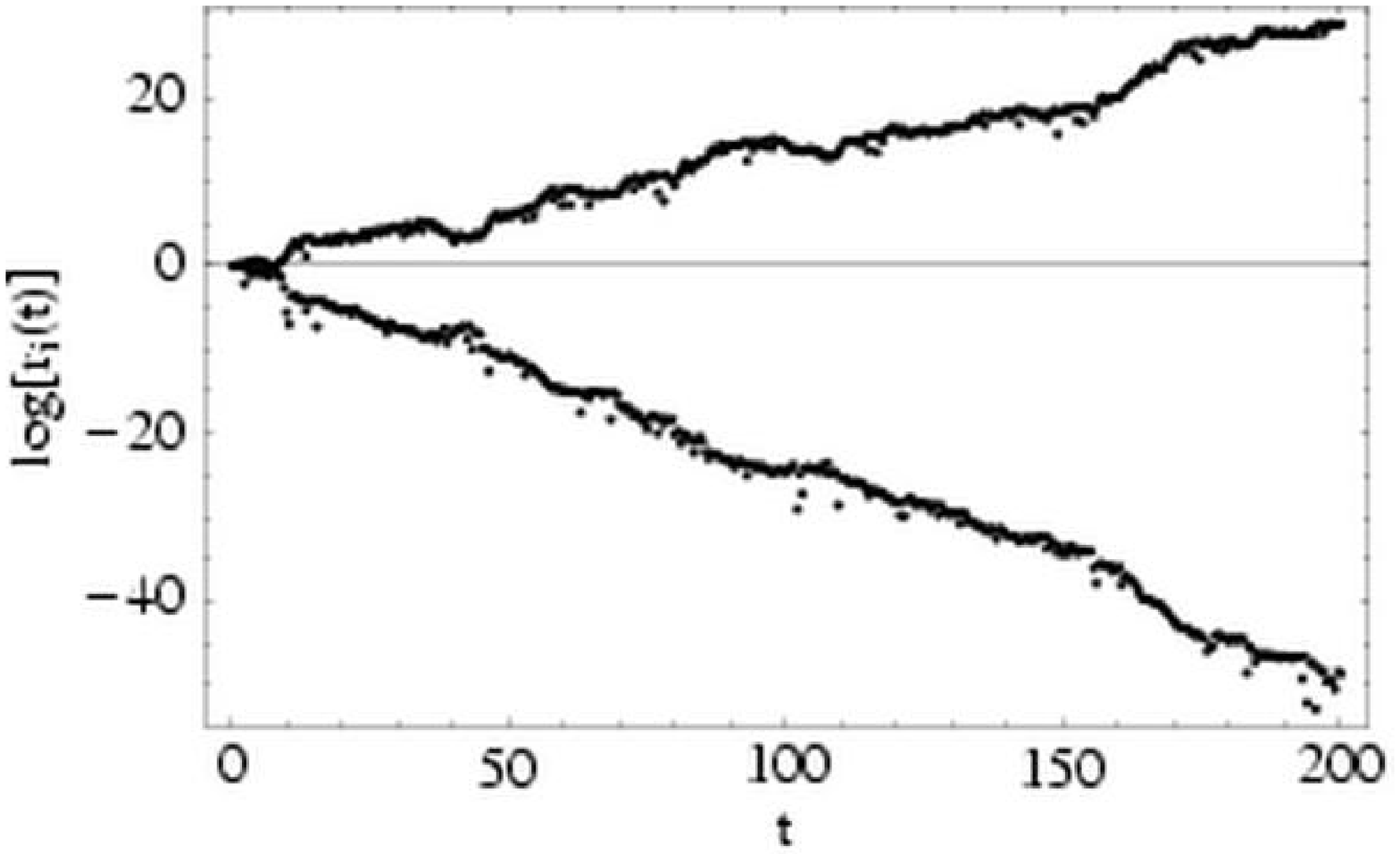}
\end{center}
\caption{\label{fig:restricted_jacobian_both}
The natural logarithms of both ellipsoid 
axes for the constrained forced damped pendulum, 
calculated using the restricted Jacobian method
(Sec.~\ref{sec:restricted_Jacobian_method}).
The slopes are the Lyapunov exponents. The results agree well
with the unconstrained case
(Fig.~\ref{fig:Lyapunov_fdp} and Table~\ref{table:compare}).}
\end{figure}

If we find $\partial y/\partial x$ using $y=\pm\sqrt{1-x^2}$, we have exactly
the same sign ambiguity problem that we had in trying to eliminate the~$y$ 
degree of freedom in the equations of motion.  The difference here is the we
need only the \emph{derivative} of $y$, not an explicit solution for $y$ in
terms of $x$, and this we can achieve by differentiating the constraint:
\begin{equation}
0 = \frac{\partial}{\partial x}(x^2 + y^2)  = 2x + 2y\frac{\partial y}{\partial
x} \Rightarrow \frac{\partial y}{\partial x} = -\frac{x}{y}.
\end{equation}
If we integrate the equations of motion using the variables~$(x, y, \omega)$,
then we have the value of $y$ at any particular time, and we never need deal
with the sign ambiguity.  Using the same trick to calculate
$\partial\dot{\omega}/\partial x$, we can write the \emph{restricted Jacobian}
as
\begin{equation}
\label{eq:fdp_restricted_Jacobian}
\mathbf{Df} = \left(
\begin{array}{cc}
	\omega\,\displaystyle{\frac{x}{y}}~&~-y\medskip\\
	\displaystyle{\frac{x}{y}} & -c\\
\end{array}
\right)
\end{equation}
We now proceed exactly as in the unconstrained Jacobian method, using the
restricted Jacobian to calculate the evolution of the initial tangent space ball.
Since we deal only with a number of coordinates equal to the true 
number of degrees
of freedom, the constraints are not a consideration, and we can
reorthogonalize exactly as before. 

The general case is virtually the same.  For $n$~coordinates
and $d$~degrees of freedom, there must be~$m=n-d$ constraint equations of the
form 
\begin{equation}
C_k(\mathbf{y}) = 0
\end{equation}
for $k = 1\ldots m$.  We must choose which $d$
coordinates to keep in the Jacobian matrix, eliminating~$m$ coordinates in the
process.  By differentiating the constraints, we arrive at $m$~\emph{linear}
equations for the derivatives of the $m$~eliminated coordinates in terms of
the~$n$ variables:
\begin{equation}
\label{eq:constraint_derivs}
\frac{\partial C_k}{\partial y_j} = 0,
\end{equation}
where~$j$ ranges over the indices of the eliminated coordinates ($j=2$,
corresponding to $y$, for the FDP).  Since these are linear equations, they are
both easy to solve and do not suffer from any sign or branch ambiguities. The
$d\times d$ Jacobian matrices that result allow the calculation of Lyapunov
exponents with all the robustness of the Jacobian method for unconstrained
systems.

\begin{figure}
\begin{center}
\includegraphics[width=3in]{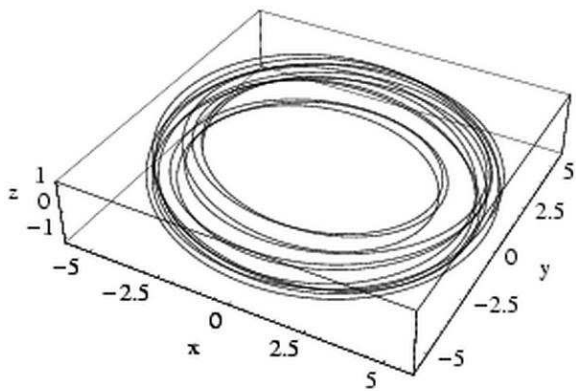}
\end{center}
\caption{
\label{fig:PN_orbit}
The orbit of a spinning relativistic binary,
calculated
using the post-Newtonian equations of motion.  The equations model \emph{two}
spinning bodies, but we use an effective one-body approach to reduce the
dynamics to the motion of one body.  Distances are measured in terms of
$GM/c^2$, where $M = m_1+m_2$ is the total mass of the system.  For a
pair of black holes, each with 10~times the mass of the Sun, 
the length unit is
$GM/c^2 = 20\,GM_\odot/c^2 = 30\,\mathrm{km}$.}
\end{figure}

\begin{figure}
\begin{center}
\includegraphics[width=3in]{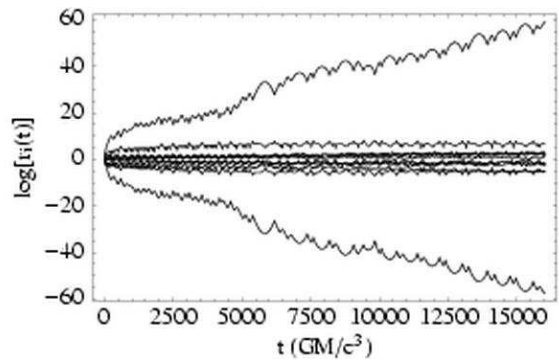}
\end{center}
\caption{
\label{fig:PN_Lyapunov}
The natural logarithms of the ellipsoid 
axes~$r_i$ vs.~$t$ for the system shown in Fig.~\ref{fig:PN_orbit}. 
Time is measured in units of $GM/c^3$, where $M = m_1 + m_2$ is the total mass
of the system. For two
10~solar-mass black holes, the time unit is $GM/c^3 = 20\,GM_\odot/c^3 = 
10^{-4}~\mathrm{s}$.
The spin magnitudes are fixed, so that
each spin vector represents only two true degrees of freedom.  
We deal with this
constraint by using the restricted Jacobian method
(Sec.~\ref{sec:restricted_Jacobian_method}).
Two nonzero exponents are clearly visible, but all the others are consistent
with zero. Note the~$\pm\lambda$ symmetry characteristic of Hamiltonian 
systems.}
\end{figure}

We considered the constrained forced damped pendulum for purposes of
illustration, but it is admittedly artificial.  A more realistic example is
shown in Fig.~\ref{fig:PN_orbit}, which illustrates the dynamics of two
spinning black holes with comparable masses.  (Such systems are of considerable
interest for ground-based gravitational wave detectors such as the LIGO
project.)  The equations of motion come from the Post-Newtonian (PN) expansion
of full general relativity---essentially, a series expansion in the
dimensionless velocity~$v/c$, where the first term is ordinary Newtonian
gravity and the higher-order terms are post-Newtonian corrections (see, e.g.,
\cite{Damour2001,DJS2000,DS1988}).  The
constraint comes from the spins of the black holes: it is most natural to think
of the spin as having \emph{two} degrees of freedom (a fixed magnitude with two
variable angles specifying the location on a sphere), but the equations of
motion use all \emph{three} components of each hole's spin.  We apply the
methods described  above to eliminate one of the spin degrees of freedom for
each black hole, using the constraints
\begin{equation}
S_{x,i}^2 + S_{y,i}^2 + S_{z,i}^2 = S_i^2 = \mbox{const.},\qquad i \in \{1, 2\}.
\end{equation}
Using the \emph{effective one-body} approach~\cite{Damour2001}, 
\emph{a priori} the system has 12
degrees of freedom: three each for relative position~$\mathbf{x}$,
momentum~$\mathbf{p}$, and the spins $\mathbf{S}_1$ and~$\mathbf{S}_2$.
Eliminating two spin components leaves 10~true degrees of freedom.  As a
result, the system has 10 Lyapunov exponents, as shown in
Fig.~\ref{fig:PN_Lyapunov}; note in particular the~$\pm\lambda$ symmetry
characteristic of Hamiltonian systems.

\subsection{Constrained ellipsoid method}
\label{sec:constrained_ellipsoid_method}

\begin{figure}
\begin{center}
\includegraphics[width=3in]{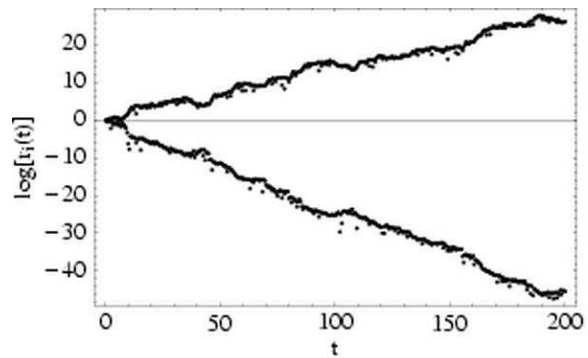}
\end{center}
\caption{\label{fig:constrained_jacobian_both}
The natural logarithms of both ellipsoid 
axes~$r_i$ vs.~$t$ for the constrained forced damped pendulum, 
calculated using the constrained ellipsoid method
(Sec.~\ref{sec:constrained_ellipsoid_method}).
The slopes are the Lyapunov exponents. The results agree well
with the unconstrained case
(Fig.~\ref{fig:Lyapunov_fdp} and Table~\ref{table:compare}).}
\end{figure}

The restricted Jacobian method relies on eliminating spurious degrees of
freedom from the Jacobian matrix, but such a prescription relies on making a
choice---namely, which coordinates to eliminate.  Each choice results in a
different Jacobian matrix.  Since calculating the Jacobian matrix even once can
be a formidable task for sufficiently complicated systems, it is valuable to
have a method that uses the \emph{full} Jacobian---treating all coordinates as
independent---which can be calculated once and then never touched again.  This
requirement leads to the \emph{constrained ellipsoid method}, which uses the
full Jacobian matrix to evolve \emph{constraint-satisfying} tangent vectors, 
collectively 
referred to as a ``constrained ellipsoid.''  When recording ellipsoid
axis growth, we extract from each vector a number of components equal to the
true number of degrees of freedom, resulting in vectors that can be
orthogonalized and (if necessary) normalized just as in the unconstrained case.

A detailed description of the constrained ellipsoid algorithm appears below,
but we first present an important prerequisite: calculating
constraint-satisfying  tangent vectors.  Let a tilde denote a vector with
dimension~$d$ equal to the true number of degrees of freedom (as in
Sec.~\ref{sec:complications}). We construct a full tangent vector~$\bm{\xi}$
(with $n$~components) from a $d$-dimensional
vector~$\tilde{\bm{\xi}}$ at a point~$\mathbf{y}$ on the flow as follows:
\begin{enumerate}

    \item Let $\tilde{\mathbf{y}}' = \tilde{\mathbf{y}} +
    \epsilon\tilde{\bm{\xi}}$ for a suitable choice of~$\epsilon$.

    \item Fill in the missing components of~$\tilde{\mathbf{y}}'$ using the
    constraints to form~$\mathbf{y}'$ as in Sec.~\ref{sec:complications}.

	\item Infer the full tangent vector~$\bm{\xi}$ using 
\begin{equation}
\label{eq:infer}
    \bm{\xi} =
\frac{\mathbf{y}' - \mathbf{y}}{\epsilon}.
\end{equation}    

\end{enumerate}
Setting the initial conditions is now simple: form a random~$d\times d$ matrix,
orthonormalize it, and then infer the full $d\times n$ matrix using the method
above on each column.  
The construction of constraint-satisfying tangent vectors described 
above is also
necessary in the reorthogonalization steps of the constrained ellipsoid method.

The full method is an adaptation of the Jacobian method from
Sec.~\ref{sec:detail}:
\begin{enumerate}

    \item  Construct a random $d\times d$ matrix and orthonormalize it to form
    a unit ball.  Use the constraints to infer the full $d\times n$
    matrix~$\mathbf{U}$.

	\item Evolve the system forward using the equations of motion and the
    evolution equation for~$\mathbf{U}$,
\begin{equation}
\dot{\mathbf{U}} = \mathbf{Df}\cdot\mathbf{U}.
\end{equation}

	\item At each time~$T$, extract the relevant eight components from each
	tangent vector to form a $d\times d$~ellipsoid, orthonormalize it, and then
	fill in the missing components using the constraints, yielding again a
	$d\times n$~matrix.  The restricted norms of the $d$~tangent vectors
    contribute to the running sum for the logs of the ellipsoid axes
    [Eq.~(\ref{eq:running_sum})].

\end{enumerate}
It is important to note that, unlike the other Jacobian methods, rescaling every
time
time~$T$ (or some similar method) is required for the inference equation
[Eq.~(\ref{eq:infer})], since the product of~$\epsilon$ and the components of
$\bm{\xi}$ must be small for the inference to work correctly.  
The method only works if the system is renormalized 
regularly, so the value of~$T$ should be chosen to be small enough that no
principal ellipsoid axis grows too large.

As before, we use the constrained FDP model for purposes of illustration.
Treating each coordinate as independent yields [upon differentiation of
Eq. (\ref{eq:fdpCartesian})]:
\begin{equation}
\label{eq:fdp_constrained_Jacobian}
\mathbf{Df} = \left(
\begin{array}{ccc}
	\frac{\partial\dot{x}}{\partial x} & \frac{\partial\dot{x}}{\partial y}
         & \frac{\partial\dot{x}}{\partial\omega}\\
	\frac{\partial\dot{y}}{\partial x} & \frac{\partial\dot{y}}{\partial y} 
        & \frac{\partial\dot{y}}{\partial\omega}\\
    \frac{\partial\dot{\omega}}{\partial x} & 
            \frac{\partial\dot{\omega}}{\partial y} 
            & \frac{\partial\dot{\omega}}{\partial\omega}\\
\end{array}
\right)
= \left(
\begin{array}{ccc}
	0 &~-\omega~~ & -y\\
	\omega &~~0~~& x\\
	0 &~-1~~& -c\\
\end{array}
\right)
\end{equation}
The coordinates are not independent, of course, but this Jacobian matrix
satisfies Eq.~(\ref{eq:fydy}) as long as the deviation is
constraint-satisfying.  For example, using the full deviation vector
$\delta\mathbf{y} = (\delta x, \delta y, \delta\omega)$ with
Eq.~(\ref{eq:fdp_constrained_Jacobian}) 
gives the same result as the restricted deviation vector
$\delta\tilde{\mathbf{y}} = (\delta x, \delta\omega)$ with
Eq.~(\ref{eq:fdp_restricted_Jacobian}), as long as
$\delta y = -(x/y)\,\delta x$.  As a result, the Lyapunov exponents calculated
with the constrained ellipsoid method
(Fig.~\ref{fig:constrained_jacobian_both}) agree closely with the
restricted Jacobian method (and with the original unconstrained results
[Fig.~(\ref{fig:Lyapunov_fdp})]).

\begin{figure*}
\begin{center}
\begin{tabular}{ccc}
\includegraphics[width=2in]{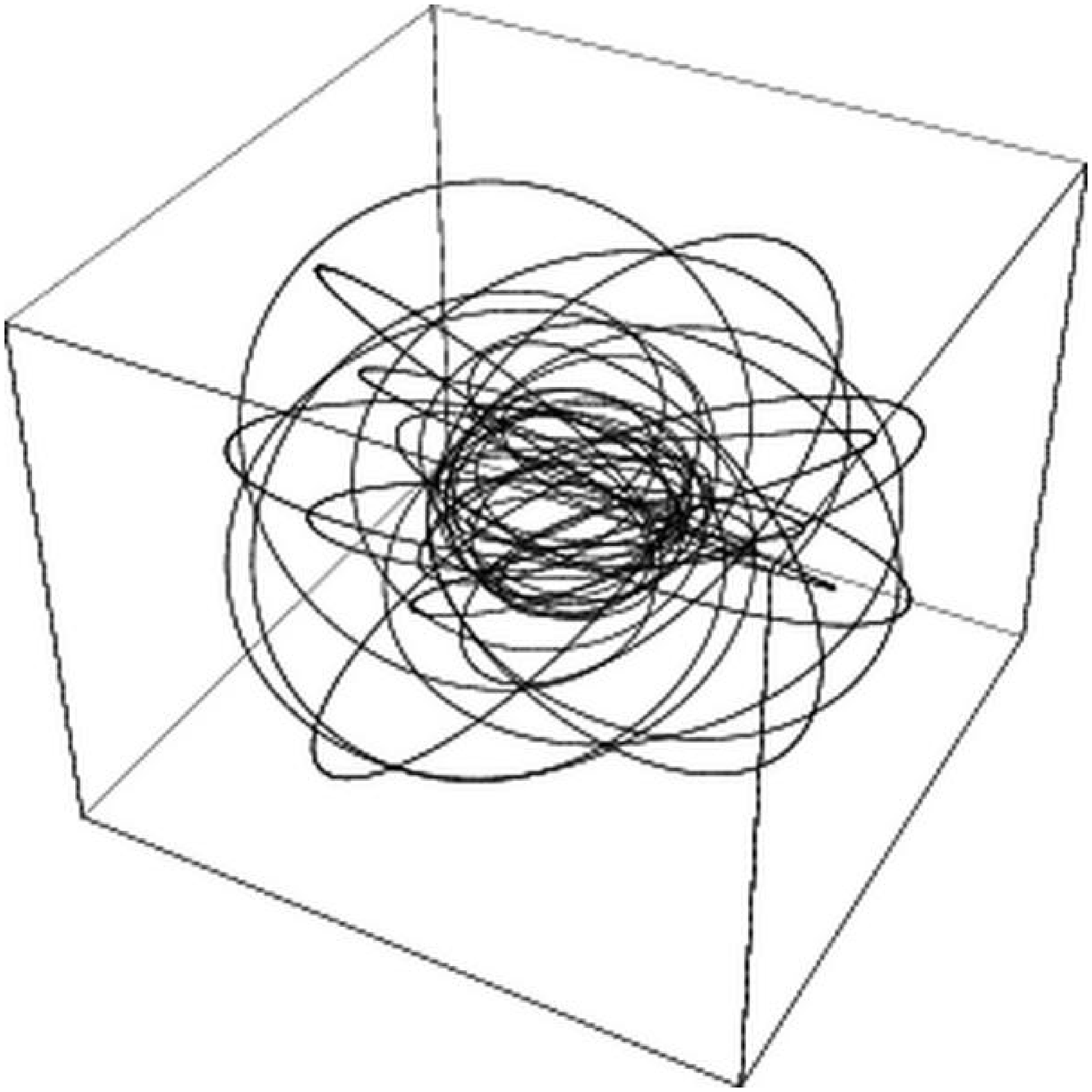} &
\medskip &
\includegraphics[width=2in]{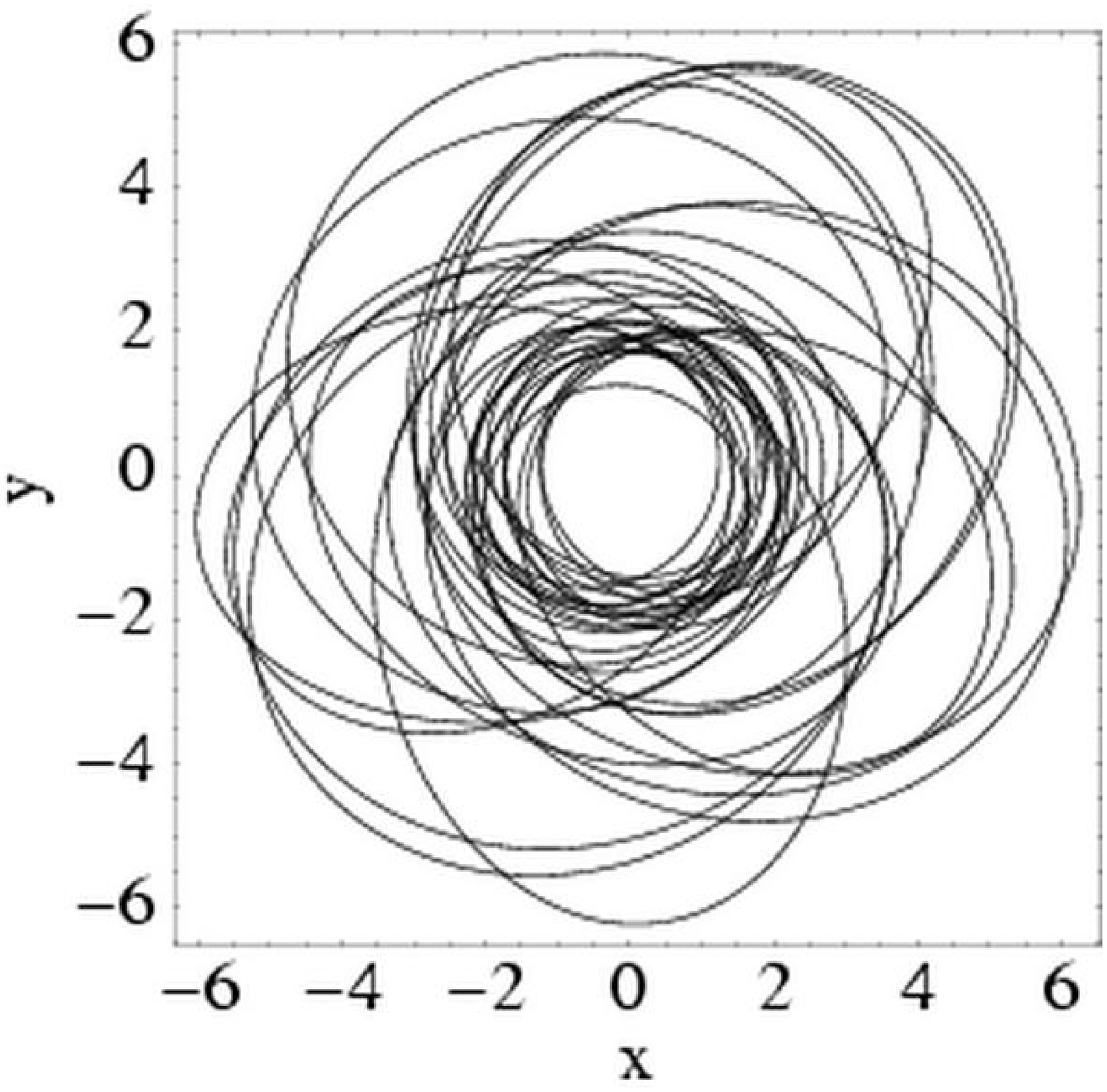}
\end{tabular}
\end{center}
\caption{
\label{fig:Papapetrou_orbit}
The orbit of a small spinning
compact object (such as a solar-mass black hole) 
in the spacetime of a rotating supermassive black hole.
(a) The orbit embedded in spherical polar coordinates;
(b) the orbit's projection onto the $x$-$y$ plane.  The lengths are expressed in
terms of~$GM/c^2$, where~$M$ is the mass of the central black hole.  For a
maximally spinning black hole, the horizon radius is $r_H = GM/c^2$.  For the
supermassive black hole at the center of the Milky Way, 
$M =3\times10^6\,M_\odot$~\cite{MilkyWay}, 
which corresponds to a length unit of $GM/c^2 = 4.4\times10^9\,\mathrm{m}$.
The system shown here is chaotic
(Fig.~\ref{fig:Papapetrou_Lyapunov}), although this orbit represents a limiting
case of the equations that is not physically realizable~\cite{Hartl_2002_1}.}
\end{figure*}

\begin{figure}
\begin{center}
\includegraphics[width=3in]{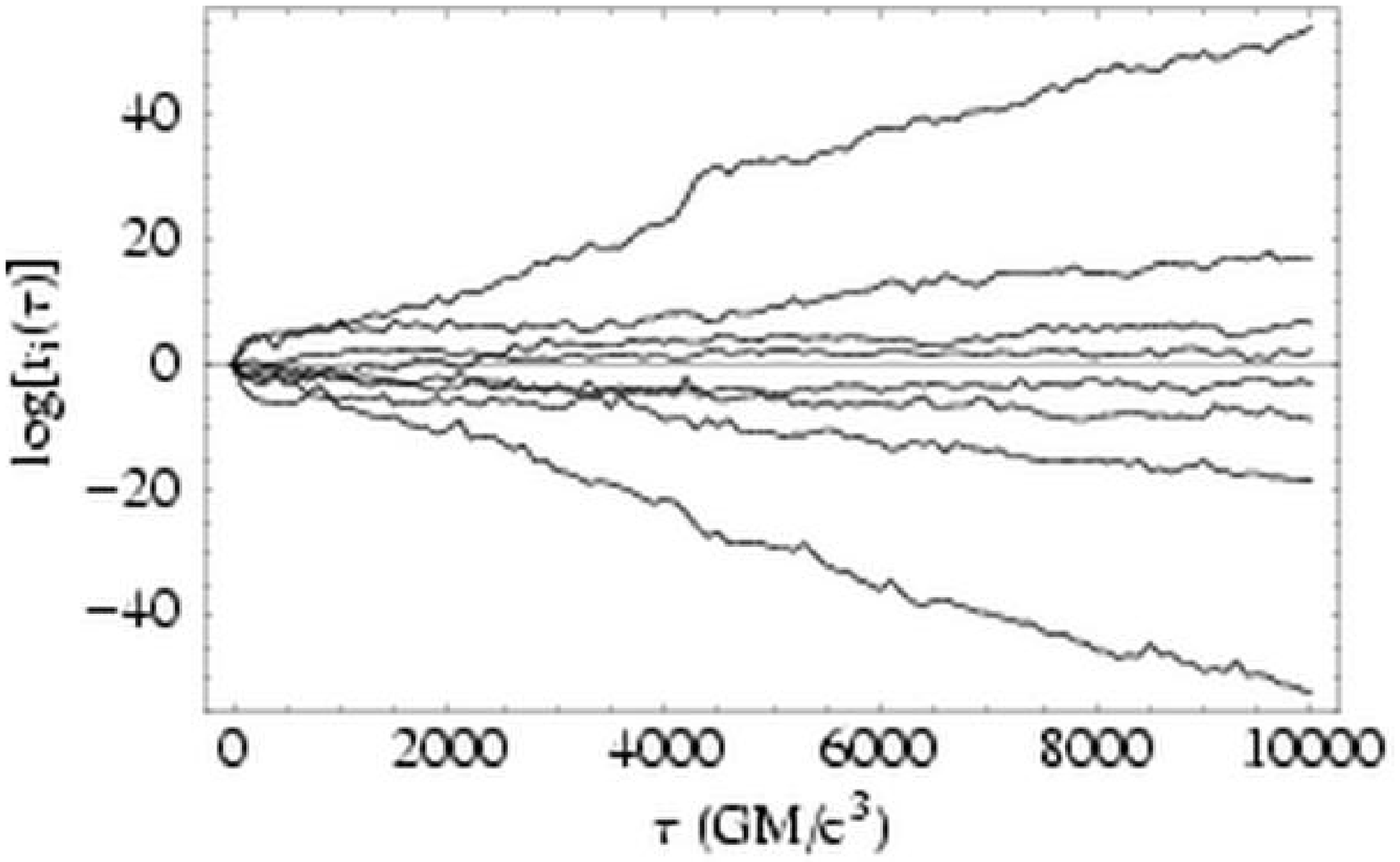}
\end{center}
\caption{
\label{fig:Papapetrou_Lyapunov}
The natural logarithms of the ellipsoid axes for the system shown in
Fig.~\ref{fig:Papapetrou_orbit} vs.~relativistic proper time~$\tau$, in units
of~$GM/c^3$, where~$M$ is the black hole's mass.  For the
supermassive black hole at the center of the Milky Way, 
$M =3\times10^6\,M_\odot$~\cite{MilkyWay}, 
which corresponds to a time unit of $GM/c^3 = 15\,\mathrm{s}$.  The largest
Lyapunov exponent is 
$\lambda_\mathrm{max}\approx5\times10^{-3}\,(GM/c^3)^{-1}$, which corresponds to
an $e$-folding timescale of $\tau_\lambda = 1/\lambda =
2\times10^2\,GM/c^3$.  For $M = 3\times10^6\,M_\odot$, this means that
nearby trajectories diverge by a factor of~$e$ in the local (Lorentz) frame of
an observer on this orbit
in a time $\tau = 3000\,\mathrm{s} = 50\,\mathrm{min.}$
We find nonzero
exponents in this system only for physically unrealistic values of the small
body's spin
(\cite{Hartl_2002_2}).}
\end{figure}

As a final example of the constrained ellipsoid method, consider
Fig.~\ref{fig:Papapetrou_orbit}, which shows a solution to equations that model
a relativistic spinning test particle (e.g., a black hole or neutron star)
orbiting a supermassive rotating black hole.   (The case illustrated is a
limiting case of the equations, which is mathematically valid but not
physically realizable; see~\cite{Hartl_2002_1}.)  These equations (usually
called the Papapetrou equations) are highly constrained, so a na\"{\i}ve
calculation of the Lyapunov exponents is not correct.  It was the complicated
nature of the Jacobian matrix for this system that originally motivated the
development of the methods in this section~\cite{Hartl_2002_1}.  A Lyapunov
plot corresponding to the orbit in Fig.~\ref{fig:Papapetrou_orbit} is shown in
Fig.~\ref{fig:Papapetrou_Lyapunov}. Note especially the $\pm\lambda$ symmetry,
a result of the Hamiltonian nature of the equations of motion. 

\section{Comparing the methods}
\label{sec:comparing}

\begin{figure}
\begin{center}
\includegraphics[width=3in]{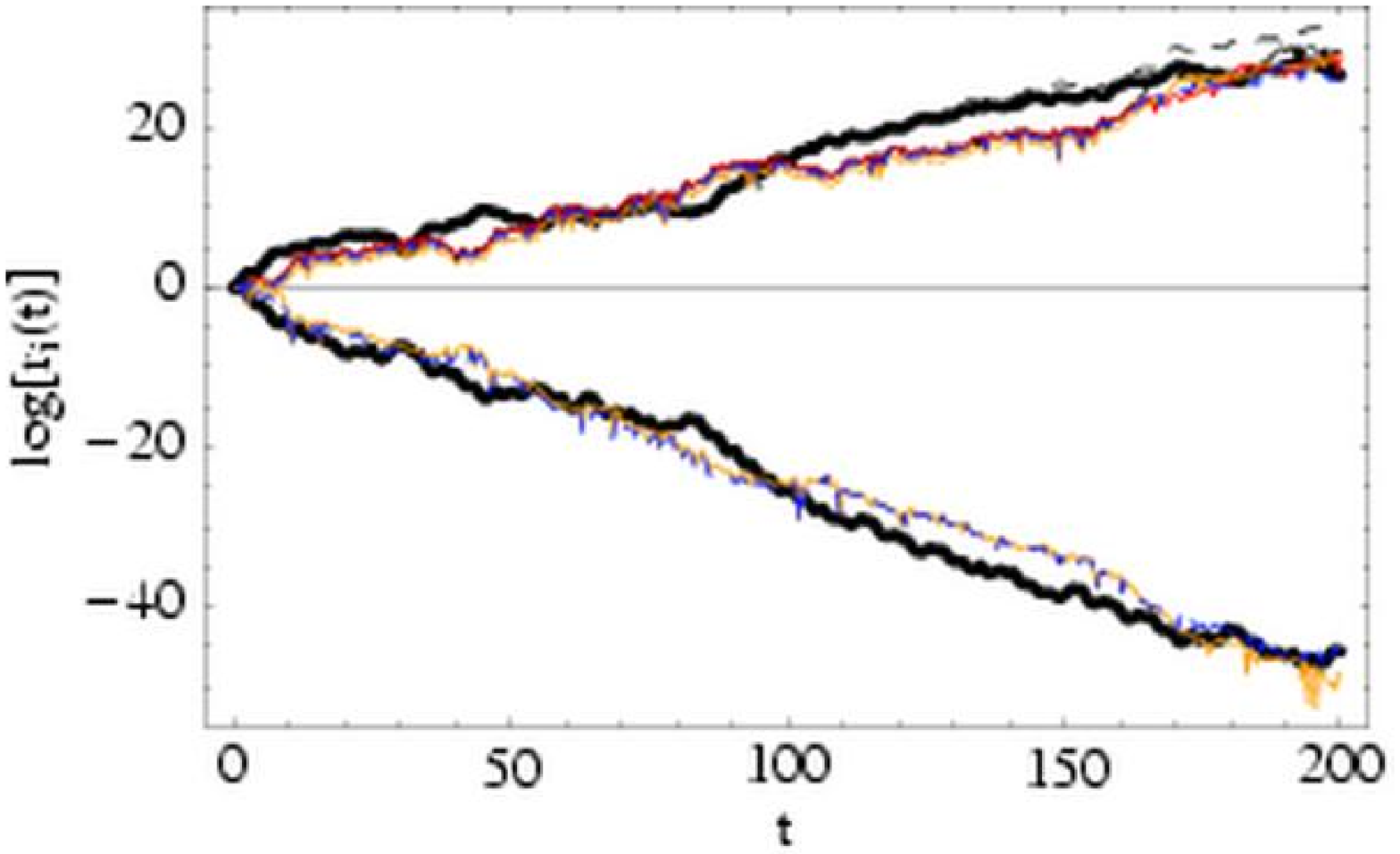}
\end{center}
\caption{
\label{fig:superposition}
Natural logarithms of the ellipsoid axes vs.~$t$ for the unconstrained deviation
vector method (dashed), the unconstrained Jacobian
method from Fig.~\ref{fig:Lyapunov_fdp} (thick) and \emph{all} the constrained methods.  
The constrained methods include the
following: rescaled deviation vector (black), Jacobian with 
single constraint-satisfying tangent vector (red), restricted Jacobian (orange),
and constrained ellipsoid (dashed blue).  
(The colors appear as shades of gray in print
versions of this paper.) 
All the constrained methods start with exactly the same
initial conditions.}
\end{figure}

A summary plot of all the methods discussed in this paper, applied to the
forced damped pendulum, appears in Fig.~\ref{fig:superposition}.  It is evident
that all the methods agree closely.  A more quantitative comparison appears in
Table~\ref{table:compare}, which gives error estimates based on integrations
using fixed initial conditions and random initial deviations.  This table was
produced by using an initial point produced from the \emph{final} values of a
previous long integration, which avoids any transient effects due to starting
at a point not on the attractor.  The estimates for the exponents use a final
time of $t_f = 10^4$, with 100~randomly chosen values for the deviation vector
or initial ball.  All the methods agree on the mean exponents to within one
standard deviation of the mean.   (Recall that we omit the zero exponent
associated with the time ``degree of freedom.'')

\begin{table*}
\caption{\label{table:compare}
Comparison of different Lyapunov exponent methods applied to the forced damped
pendulum.  We consider both the unconstrained [Eq.~(\ref{eq:fdp})] and 
constrained [Eq.~(\ref{eq:fdpCartesian})] formulations.  The integrations have 
a final time~$t_f=10^4$, and for each method
we consider 100 random initial deviations.   
We calculate the positive
exponent ($\lambda_1$) and, if possible, the negative exponent ($\lambda_3$) as
well.  (We omit the zero exponent ($\lambda_2$) for brevity.)
The error
estimates are the standard deviations in the mean, $\sigma/\sqrt{N}$.
The deviation
vector methods are all rescaled.  The constrained ellipsoid method rescales and
reorthogonalizes every time~$T=1$, and uses a value of~$\epsilon=10^{-6}$ for
the tangent-vector inference [Eq.~(\ref{eq:infer})].
The error goal is a fractional error
of~$10^{-10}$ per step. 
}
\medskip
\begin{center}
\begin{tabular}{|c|c|c|}\hline
Method & $\lambda_1$ & $\lambda_3$ \\ \hline
unconstrained~deviation~vector & $0.1610\pm0.00050$ &  \\
unconstrained~Jacobian & $0.1608\pm0.00050$ & $-0.2618\pm0.00053$ \\
constrained~deviation vector & $0.1608\pm0.00051$ &  \\
constrained~Jac. (1~tangent vector) & $0.1605\pm0.00048$ &  \\
restricted Jacobian & $0.1607\pm0.00048$ & $-0.2614\pm0.00055$ \\
constrained~ellipsoid & $0.1605\pm0.00050$ & $-0.2617\pm0.00051$ \\ \hline
\end{tabular}
\end{center}
\end{table*}

\subsection{Speed}

The various methods for calculating the exponents differ significantly in their
execution time, as shown in Table~\ref{table:speed}.  Generally speaking, the
deviation methods are faster than their Jacobian method counterparts, which is
no surprise---the deviation vector methods involve fewer differential
equations. More surprising is the performance penalty for the restricted
Jacobian method.  This is the result of a significantly smaller typical
step-size in the adaptive integrator needed to achieve a particular error
tolerance.  The restricted Jacobian may result in a system of equations that is
more difficult to integrate because of the elimination of simple degrees of
freedom with the potentially complicated solutions to the  constraint
derivative equations $\partial{C_k}/\partial{y_i} = 0$
[Eq.~(\ref{eq:constraint_derivs})]. On the other hand, the performance penalty
of the restricted Jacobian method is probably worth the gain in robustness, as
discussed below.  Moreover, for other systems (e.g., the system shown in
Figs.~\ref{fig:PN_orbit} and~\ref{fig:PN_Lyapunov}), the restricted Jacobian
method is comparable in speed to the other Jacobian methods.

\begin{table}
\caption{\label{table:speed}
Timing comparison for
different Lyapunov exponent methods applied to the forced damped
pendulum.  The times (on a 2~GHz Pentium~4) for a final time of~$t_f=10^4$
are in seconds: $t_1$ for the positive exponent~$\lambda_1$ and
$t_{1\mathrm{-}3}$
for the negative exponent~$\lambda_2$; 
we omit the zero exponent ($\lambda_2$) for brevity. 
(We write $1\mathrm{-}3$ to emphasize that
calculating $\lambda_3$ also calculates $\lambda_1$ as a side-effect.) 
We consider both the unconstrained [Eq.~(\ref{eq:fdp})] and 
constrained [Eq.~(\ref{eq:fdpCartesian})] formulations.  
The integrations use a C++ Bulirsch-Stoer integrator adapted
from~\cite{NumRec}.  The deviation
vector methods are rescaled, and the constrained ellipsoid method rescales and
reorthogonalizes every time~$T=1$.  The error goal is a fractional error
of~$10^{-10}$ per step.  The relatively small difference between deviation
vector and Jacobian methods is the result of the small number of degrees of
freedom; for larger systems (with larger Jacobians) the difference can become
large~\cite{Hartl_2002_2}. We note that the 
restricted Jacobian method is unusually slow for
the forced damped pendulum, but this is not generally the case.  
}
\medskip
\begin{center}
\begin{tabular}{|c|c|c|}\hline
Method & $t_1$ & $t_{1\mathrm{-}3}$ \\ \hline
unconstrained~deviation~vector & $2.57$ &  \\
unconstrained~Jacobian & $3.65$ & $5.16$ \\
constrained~deviation vector & $3.51$ &  \\
constrained~Jacobian (1~tangent vector) & $4.05$ &  \\
restricted Jacobian & $35.3$ & $45.0$ \\
constrained~ellipsoid & $4.30$ & $5.88$ \\ \hline
\end{tabular}
\end{center}
\end{table}

\subsection{Robustness}
\label{sec:robustness}

Numerical methods are more useful if they are relatively insensitive to small
changes in implementation details, and the Jacobian methods win in this
category.  When reorthogonalization occurs every time step, \emph{without}
rescaling, the plain Jacobian method is virtually bulletproof.  The rescaling
in this case can even occur only when the tangent vector norms reach very large
or small values, say $\|\bm{\xi}\|\approx 10^{\pm100}$. This robustness also
applies to the restricted Jacobian method, which is considerably less finicky
than any other method for constrained systems, and we recommend its
implementation if practical.

\begin{figure}
\begin{center}
\includegraphics[width=3in]{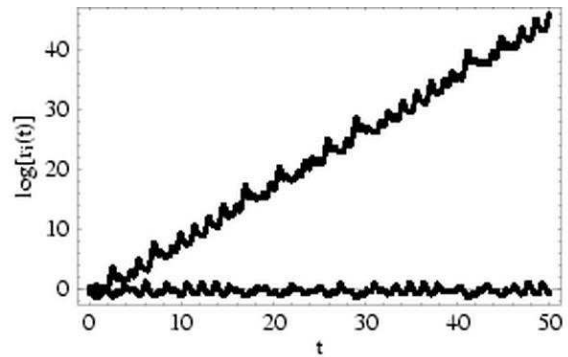}
\end{center}
\caption{
\label{fig:good_GS}
The natural logarithms of the two larger ellipsoid axes for the Lorenz system
using the Gram-Schmidt algorithm, with the axes rescaled every $T=10^{-3}$.
The largest and smallest directions differ by less than~$2\%$ when rescaling
this frequently, but the axes nevertheless converge rapidly to the correct
directions (as determined by the Jacobian method,
Fig.~\ref{fig:Lorenz_Jacobian_2}).  Numerical investigations
confirm that this robustness persists at least down to $T=10^{-5}$.
}
\end{figure}

\begin{figure}
\begin{center}
\includegraphics[width=3in]{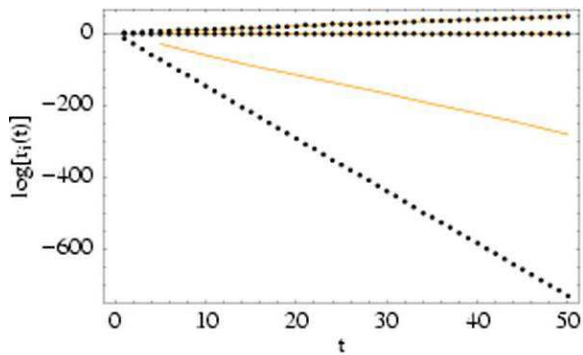}
\end{center}
\caption{
\label{fig:Lorenz_compare}
The natural logarithms of the ellipsoid axes for the Lorenz system, with
reorthogonalization/rescaling every time~$T=1$ (dark dots) and $T=5$ (light
lines).  The two larger exponents agree exactly, but the negative exponent is
incorrect due to roundoff error, since the smallest axis shrinks from unity to a
size of
$e^{-5\times14.57}\approx 2\times10^{-32}$ in a time $T=5$.
}
\end{figure}

Jacobian methods that rescale and reorthogonalize every time~$T$ are less
robust, since \emph{a priori} we have no knowledge of appropriate values
for~$T$.  Experimentation in this case is required to find good values of~$T$;
for the Lorenz system, $T=1$ works well, but $T=5$ leads to inaccurate
estimates for the negative exponent, as seen in Fig.~\ref{fig:Lorenz_compare}. 
It is better to err in the direction of small times, since the Gram-Schmidt
procedure is quite robust: even when rescaling occurs on very short
timescales---so that the longest axis has almost no chance to outgrow
the other principal axes---the Gram-Schmidt method still converges to the
correct exponents (Fig.~\ref{fig:good_GS}).  Using the Gram-Schmidt algorithm
to find the principal axes benefits from a strong feedback mechanism, insuring
rapid convergence to the correct axes.  Using a very small value for~$T$
greatly increases the execution time, of course.  A useful prescription in
practice is to do a short integration with~$T$ chosen to be small compared to
any characteristic timescales in the problem, in order to obtain a first
estimate for the exponents.  We may then choose~$T$ to be as large as we like, 
consistent with the avoidance of unacceptable roundoff error.  

The constrained ellipsoid method  is dependent on frequent rescaling to keep
the size of the tangent vectors small, since the inference scheme represented
by Eq.~(\ref{eq:infer}) fails for large vector norms.  As a result, this method
suffers from the complexity of all time~$T$ methods, i.e., it requires care in
choosing an appropriate value of~$T$.  In addition, the value of $\epsilon$ in
Eq.~(\ref{eq:infer}) must be chosen carefully to achieve accurate
tangent-vector inferences: the method relies on small values of $\epsilon$ for
accuracy, but values that are too small suffer from roundoff errors.  
It is advisable to calibrate the value of $\epsilon$ so that the largest
Lyapunov exponent agrees with the result of a second method (such as the single
tangent-vector method or the deviation vector method), as discussed
in~\cite{Hartl_2002_1}.  Such a calibration was required to produce the values
in Table~\ref{table:compare}; the largest exponent calculated using the
constrained ellipsoid method differs from the other methods by several standard
deviations when using $\epsilon=10^{-5}$ for the inference, but agrees well
when using $\epsilon=10^{-6}$.

\begin{figure}
\begin{center}
\includegraphics[width=3in]{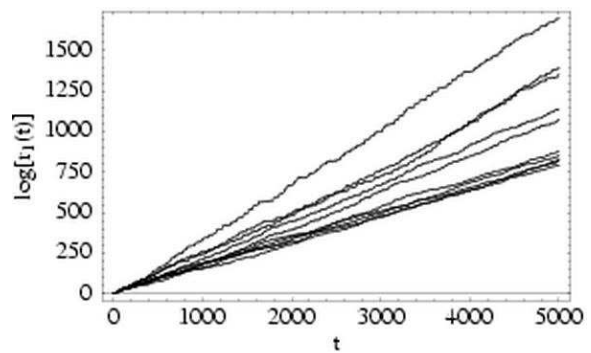}
\end{center}
\caption{
\label{fig:deviation_compare_epsilon}
The natural logarithms of the largest ellipsoid axis for the constrained 
forced damped
pendulum, calculated using the rescaled deviation vector method for varying
sizes of the initial deviation.
We vary the size of the initial deviation vector from $\epsilon_0=10^{-4}$
(bottom) to $\epsilon_0=10^{-13}$ (top).  Values of $\epsilon_0$ between 
$10^{-4}$
and $10^{-8}$ agree closely, but smaller values lead to
erroneously high values for the Lyapunov exponent.  It is important to calibrate
the deviation vector method using the Jacobian method
(Sec.~\ref{sec:Jacobian_method}) if possible.
}
\end{figure}

Finally, the deviation vector methods are all very fast, but they are sensitive
to the size~$\epsilon_0$ of the initial deviation vector.  The rescaled methods
are particularly inaccurate if the value of $\epsilon_0$ is too small, which
leads to roundoff error in the initial size of the deviation vector and can
give inaccurate results, as shown in Fig.~\ref{fig:deviation_compare_epsilon}. 
These methods should be used with care, and should always be double-checked
with a Jacobian method if possible.

\section{Summary and conclusion}

Chaotic solutions exist for an enormous variety of nonlinear dynamical
systems.  Lyapunov exponents provide an important quantitative measure of this
chaos.  We have presented a variety of different methods for calculating these
exponents numerically, both for constrained and unconstrained systems.  Both
types of systems can be investigated using deviation vector methods or Jacobian
methods.  Deviation vector methods use the equations of motion to evolve two
nearby trajectories in phase space to determine the time-evolution of the small
deviation vector joining the trajectories.  This family of methods is
computationally fast, but yields only the largest exponents, and also suffers
from sensitivity to the size of the initial deviations. The
Jacobian methods share the use of the Jacobian matrix of the system as a
rigorous measure of the local phase-space behavior.  They are computationally
robust in general, and can be used to determine multiple exponents, but this
comes at the cost of execution speed.

Calculating Lyapunov exponents for constrained systems presents a variety of
complications, all revolving around the notion of constraint-satisfying
deviations: ``nearby'' trajectories must be chosen carefully to insure that
they satisfy the constraints.  We have presented several methods for dealing
with these complications, including a deviation vector method and two Jacobian
methods: the restricted Jacobian method, which eliminates spurious degrees of
freedom in the Jacobian by differentiating the constraints; and the constrained
ellipsoid method, which uses the full Jacobian matrix to evolve
constraint-satisfying tangent vectors.  These methods allow the determination
of all $d$~Lyapunov exponents for systems with $d$~degrees of freedom.

\section*{Acknowledgments}

Thanks to Sterl Phinney for encouragement and valuable comments.  This work was
supported in part by NASA grant NAG5-10707.


\appendix

\section{Ellipsoid axes and the singular value decomposition}

In this appendix, we discuss an alternative method for calculating the
ellipsoid axes used in the Jacobian method, namely, calculating the ellipsoid
axes exactly.  The method described seems superior on paper to the Gram-Schmidt
technique described in Sec.~\ref{sec:Jacobian_method}, but suffers from subtle
complications that make it fragile in practice.  Nevertheless, within a narrow
range of validity (specified below), calculating exact ellipsoid axes provides
valuable corroboration of the principal Jacobian method discussed above.

Recall Theorem~\ref{thm:axes} from Sec.~\ref{sec:ellipsoids}, which relates the
eigensystem of the matrix $A^T A$ to the ellipsoid spanned by the columns
of~$A$. In order to find the axes of an evolving ellipsoid,  we could apply
Theorem~\ref{thm:axes} directly, but there is a mathematically equivalent
prescription that is numerically virtually bulletproof, namely, the famous
\emph{singular value decomposition}:
\begin{theorem}
\label{thm:SVD}
Let $A$ be a nonsingular $n\times n$ matrix.  Then there exist orthonormal
$n\times n$
matrices~$U$ and~$V$, and a diagonal matrix~$S$, such that
\begin{equation}
\label{eq:SVD}
A = USV^T.
\end{equation}
This is the \emph{singular value decomposition} (SVD) of~$A$, and the
values~$s_i$ in $S=\mathrm{diag}(s_1,\ldots,s_n)$ are the \emph{singular
values}.
\end{theorem}
Since~$V$ is an orthogonal matrix, we have
$V^T = V^{-1}$, so that Eq.~(\ref{eq:SVD})
is equivalent to $AV = US$.  Geometrically, this means that the image of the
unit ball~$V$ is equal to an ellipsoid whose $i$th principal axis 
is given by $s_i$ times the
$i$th column of~$U$.  $V$~in this context is a special ball, but the image of
\emph{any} unit ball is the same unique ellipsoid.
This leads to the following theorem:
\begin{theorem}
Let $A$ be a nonsingular $n\times n$ matrix, and let~$U$ and~$S$ be the matrices
resulting from the singular value decomposition of~$A$ [Eq.~(\ref{eq:SVD})].  
Then the columns of~$A$ span an
ellipsoid whose $i$th principal axis is~$s_i\,\mathbf{u}_i$, where
$S=\mathrm{diag}(s_1,\ldots,s_n)$ and $\{\mathbf{u}_i\}_{i=1}^n$ are the 
columns of~$U$.
\end{theorem}
We thus see that the singular value decomposition is equivalent to finding the
eigensystem of $A^T A$.  (See Appendix~A in~\cite{ASY1997} for proofs of these
theorems.)

\begin{figure}
\begin{center}
\includegraphics[width=3in]{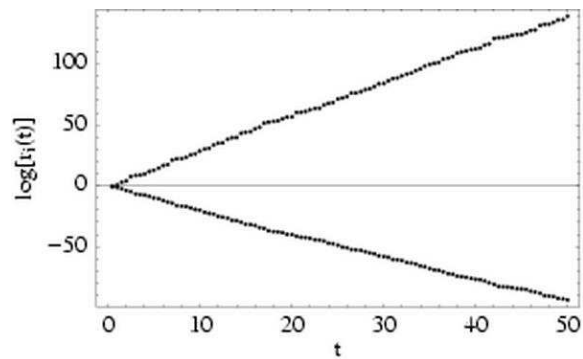}
\end{center}
\caption{
\label{fig:bad_SVD}
The natural logarithms of the two larger ellipsoid axes for the Lorenz system
using the singular value decomposition.  The axes are rescaled every $T=0.5$
to exaggerate the deviations from the correct results, but any rescaling causes
the SVD method to fail (see text).  
Compare to unrescaled SVD (Fig.~\ref{fig:good_SVD}) and
the Gram-Schmidt method with frequent rescaling (Fig.~\ref{fig:good_GS}).
}
\end{figure}

Substituting the singular value decomposition for the Gram-Schmidt procedure
leads to a replacement of step~(2) from Sec.~\ref{sec:Jacobian_method}:
\begin{itemize}

\item[($2'$)] At various times~$t_j$, replace~$\mathbf{U}$ with the orthogonal
axes of the ellipsoid defined by~$\mathbf{U}$, using the singular value
decomposition.  This can be done either every time~$T$, for some suitable
choice of~$T$, or every time the integrator takes a step. \emph{It is essential
to order the principal axes consistently.} We recommend sorting the axes so
that $s_1\geq s_2\geq\ldots\geq s_n$.

\end{itemize}

\begin{figure}
\begin{center}
\includegraphics[width=3in]{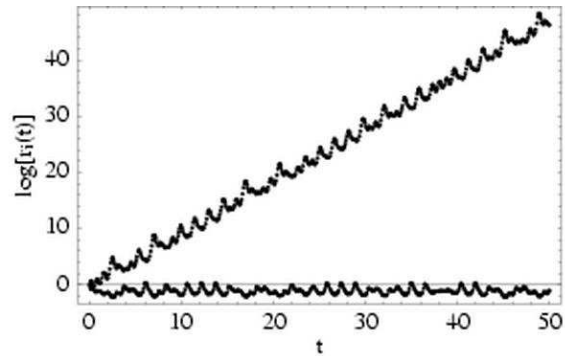}
\end{center}
\caption{
\label{fig:good_SVD}
The natural logarithms of the two larger ellipsoid axes for the Lorenz system
using the singular value decomposition, without rescaling.  The results agree
well with the Gram-Schmidt method (Fig.~\ref{fig:Lorenz_Jacobian_2}).  
Rescaling (which is always necessary if we approach
the floating point limits of $\sim\!\!10^{\pm308}$) ruins the agreement.
}
\end{figure}

Unfortunately, this prescription behaves badly when rescaling is necessary, as
shown in Fig.~\ref{fig:bad_SVD}.  The underlying cause of this is a fundamental
property of the singular value decomposition: it is only unique up to a
permutation of the ellipsoid axes.   If we adopt an ordering based on the axis
lengths, we can refer, for example, to the longest axis as axis~1.  During any
particular time period, axis~1 may grow or shrink; the only requirement is that
it be the fast-growing axis on average.  Unfortunately, rescaling the axes
causes this ordering method to fail: if axis~1 should happen to contract
between rescaling times, then the ordering based on length leads to incorrect
axis labels, since axis~1 is no longer the longest axis.  Even worse, when
ordering by axis length, the length of the longest axis is always added to the
running sum for the largest Lyapunov exponent, while the length of the smallest
axis always contributes to the smallest exponent.  This selection bias leads to
systematic errors, guaranteeing overestimates for the  absolutes values of both
the exponents (Fig.~\ref{fig:bad_SVD}).

If the system is not rescaled, there is still some initial ambiguity in axis
labels, but once axis~1 has grown sufficiently large it is very unlikely ever
to become smaller than the other axes.  Thus, after an initial expansion and
contraction phase that establishes the ordering, the axis labels remain fixed,
and the results of the (unrescaled) SVD method agree well with Gram-Schmidt
(Fig.~\ref{fig:good_SVD}).  

It should be possible in principle to follow the axis evolution by tracking the
continuous deformation of the ellipsoid.  This would mean assigning labels to
the axes and then ensuring, e.g., that axis~1 at a later time is indeed the
image of the original axis~1.  This method would require following the system
over very short timescales to guarantee the correct tracking of axes, and even
then is likely to be fragile and error-prone.  Because of these complications,
we recommend the simpler Gram-Schmidt process, which has proven to be reliable
and robust in practice.

\bibliography{mdh_lyapunov}

\end{document}